\RequirePackage{rotating}
\documentclass{tMPH2e}

\usepackage{graphicx,braket,cleveref}% Include figure files
\usepackage{rotating}
\usepackage{bm}% bold math
\usepackage[mathlines]{lineno}
\usepackage{commath,color}

\newcommand{\ce}[1]{#1}

\newcommand{\Abinitio}{\emph{Ab initio}}
\newcommand{\abinitio}{\emph{ ab initio}}

\newcommand{\X}{\mbox{X $^4\Sigma^-$}}
\newcommand{\Ap}{\mbox{A$^\prime$ $^4\Phi$}}
\newcommand{\A}{\mbox{A $^4\Pi$}}
\newcommand{\B}{\mbox{B $^4\Pi$}}
\newcommand{\C}{\mbox{C $^4\Sigma^-$}}
\newcommand{\D}{\mbox{D $^4\Delta$}}
\newcommand{\mc}{\multicolumn}
\newcommand{\Da}{\mbox{a $^2\Sigma^-$}}
\newcommand{\Db}{\mbox{b $^2\Gamma$}}
\newcommand{\Dc}{\mbox{c $^2\Delta$}}%{1$^2\Gamma$}
\newcommand{\Dd}{\mbox{d $^2\Sigma^+$}}%{$\alpha^2\Sigma^+$}
\newcommand{\De}{\mbox{e $^2\Phi$}}
\newcommand{\Df}{\mbox{f $^2\Pi$}}
\newcommand{\Dg}{\mbox{g $^2\Pi$}}

\newcommand{\cm}{cm$^{-1}$}

\newcommand{\red}[1]{{}}%\color{red} #1}}

\begin{document}

%\doi{10.1080/0026897YYxxxxxxxx}
 %\issn{1362–3028}
%\issnp{0026–8976}
%\jvol{00}
%\jnum{00} \jyear{2009} %\jmonth{10 May}

\articletype{ARTICLE}

\title{\Abinitio\ calculations to support accurate modelling of the rovibronic spectroscopy calculations of vanadium monoxide (VO)}%Electronic spectroscopy of Choice of orbitals in multi-reference configuration interaction calculations: 

\author{Laura K. McKemmish*,\footnote{* laura.mckemmish@gmail.com}
 Sergei N. Yurchenko, Jonathan Tennyson \\
Department of Physics and Astronomy, University College London, Gower Street,  London, WC1E 6BT, UK}

\maketitle

\begin{abstract}
Accurate knowledge of the rovibronic near-infrared and visible absorption spectra of transition metal diatomic species like vanadium monoxide (VO) is very important for studies of cool stellar and hot planetary atmospheres. Here, the required \abinitio\ curves are produced for the dipole moment and spin-orbit coupling, both diagonal and off-diagonal. The reliability of these curves is estimated by comparing potential energy surfaces obtained using the same methodology against experimental data (e.g. excitation energies, vibrational frequencies, bond distances). The \abinitio\ data produced here forms the basis of a new spectroscopic model for the rovibronic spectroscopy of VO. This model has been used to produce a new VO line list which considers 13 different electronic states and contains almost 640,000 energy levels and over 277 million transitions. 

Open shell transition metal diatomics are challenging species to model through \emph{ ab initio} quantum mechanics due to the large number of low-lying electronic states, significant relativistic effects (particularly strong spin-orbit coupling within and between electronic states) and strong static and dynamic electron correlation. Multi-reference configuration interaction (MRCI) methodologies using orbitals from a complete active space self-consistent-field (CASSCF) calculation are the standard technique for this kinds of system. We use different state-specific or minimal-state CASSCF orbitals for each electronic state to maximise the accuracy of the calculation. We demonstrate that this choice of orbitals significantly affects the quality of the property calculations by comparing results using CASSCF orbitals optimised for different numbers of states. 

The off-diagonal dipole moment, or the transition moment, is the critical property controlling the intensity of electronic transitions. We test the use of finite-field off-diagonal dipole moments, but found that (1) the accuracy of the excitation energies were not sufficient to allow accurate dipole moments to be evaluated and (2) computer time requirements for perpendicular transitions were prohibitive. The best off-diagonal dipole moments are calculated using wavefunctions with different CASSCF orbitals.
\end{abstract}

\begin{keywords}
ab initio, spectroscopy, MRCI, transition metal diatomic, VO
%{\bf{(Authors: Please provide three to six keywords taken from terms used in your manuscript}})
\end{keywords}\bigskip

\section{\label{sec:level1}Introduction}

\begin{figure}
\includegraphics[width=0.7\textwidth,angle=90]{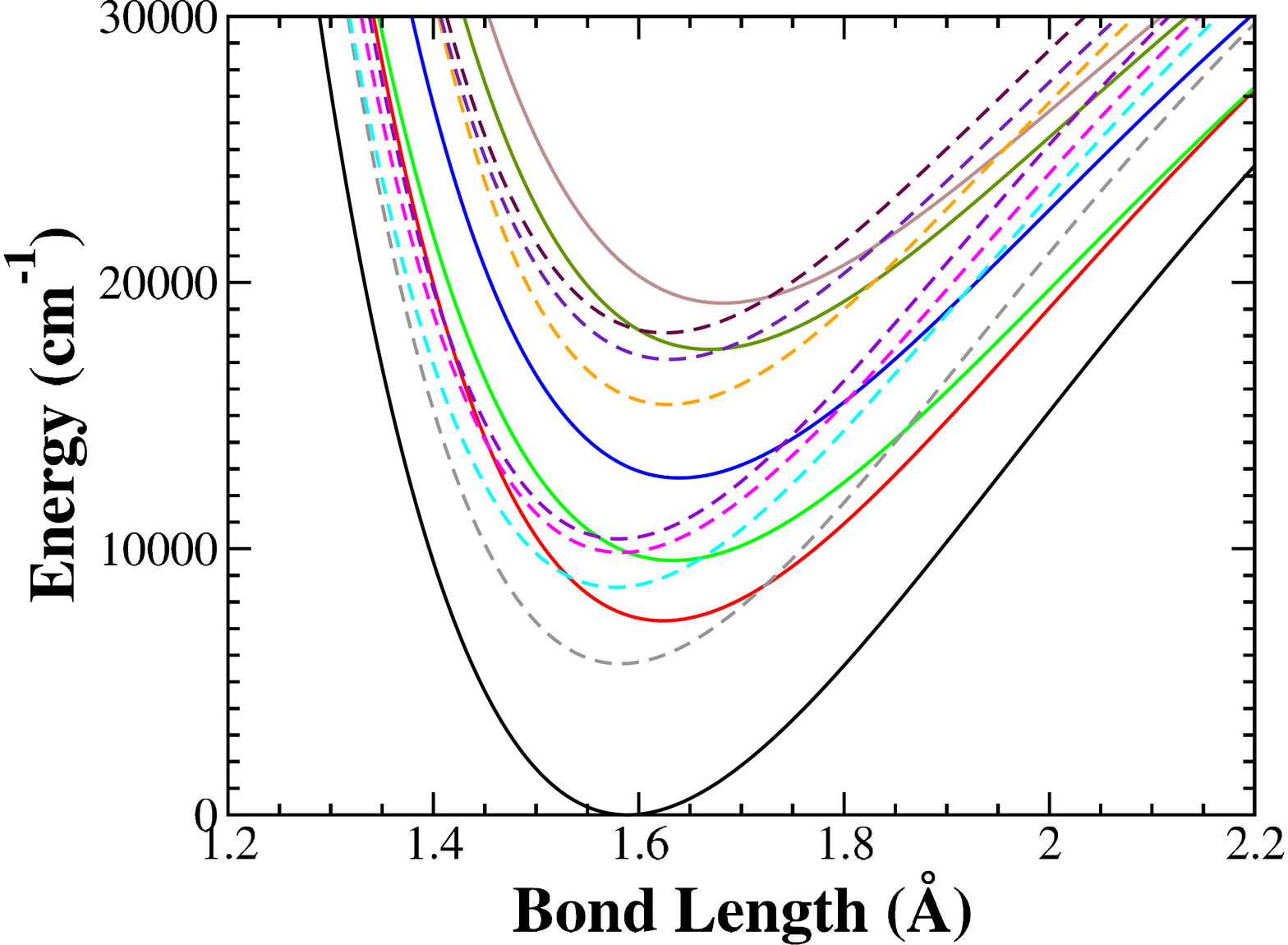}
\caption{\label{fig:PES}VO potential energy curves considered in this work. Curves in ascending order are: solid; \X, \Ap, \A, \B, \C, \D; dashed; \Da, \Db, \Dc, \Dd, \De, \Df, \Dg. }
\end{figure}

Transition metal diatomics are important absorbing species in many high
temperature systems, particularly cool stellar \cite{00CaLuPi.VO} and
hot planetary atmospheres \cite{15HoKoSn.VO}, and in industrial processes, such as biomass gasification \cite{14PuPaFl}
and the incineration stage of waste disposal \cite{02Monkhouse}, where contamination by
heavy metals, particularly mercury, is of significant environmental
concern.
Here, we focus on the vanadium oxide (VO) molecule, which is of particular interest to astronomers modelling late M-type stars and hot Jupiter exoplanets.  The potential energy curves of the low lying electronic states of VO are shown in \cref{fig:PES} for reference. Here, we report on our methodology explorations, and detail the methodology used to produce the final \abinitio\ data. % This 

A significant number of quantum chemistry calculations of VO have been
performed previously \cite{95BaMaxx.VO, 03PyWuxx.VO,03DaDeYa.VO,01BrBoxx.VO,00BrRoxx.VO,01CaSiAn.VO,08DuWaSh.VO,10KuMaxx.VO,06DuWaSh.VO}. 
For equilibrium properties and the lowest electronic state of a given spin-symmetry, single-reference methods can give good results, e.g. the coupled-cluster singles, doubles and perturbative triples, CCSD(T), calculations of Bauschlicher et al. \cite{95BaMaxx.VO}.
However, for good surfaces (particularly longer bond lengths) and other electronic states, multi-reference methods are imperative. Fortunately, %; good agreement between theory and experiment was found for the equilibrium bondlength ($r_e$), harmonic frequency ($\omega_e$), dissociation energy ($D_0$) and equilibrium dipole moment ($\mu_e$) of this state.  
computational
advances have allowed studies using the higher quality internally-contracted multi-reference
configuration interaction (icMRCI) methodology \cite{03PyWuxx.VO,07MiMaxx.VO,15HuHoHi.VO}.  
Milordos et al.  \cite{07MiMaxx.VO} performed a thorough \emph{ ab initio} study of the equilibrium properties of the nine lowest electronic states of VO using MRCI methods. 
 Milordos et al. used a
state averaged approach  to obtain results for the excited
states; however, it is unclear which states were included in each 
calculation. Total energies, equilibrium bond lengths,
electronic excitation energies, vibrational energies and equilibrium dipole moments
were given for all states, including comparison to experimental data
where available.
Equilibrium values of the spin-orbit
coupling constants are found for the triplet non-$\sigma$ states.
A 2015 study by H\"{u}bner et al.  \cite{15HuHoHi.VO} calculates the energetics of a
much larger number of electronic states. However, no dipole
moment or spin-orbit couplings were reported, except for the ground-state
equilibrium dipole moment.  H\"{u}bner et al. investigated the effect of
including $3p$ correlation on their icMRCI results. Their calculations were much more computationally demanding but did not produce significant improvement in the potential energy
curves.  In particular, H\"{u}bner et al's icMRCI calculations fail to predict the correct
ordering of the \C{} and \D{} states.

Other $3d$ transition-metal/first-row-atom diatomic species have proven to have similar challenges.\cite{00Harris.FeH,00KaMaxx.Re,04BoMaxx.Re,03TzMaxx.FeC,97Laxxxx.TiO,01Doxxxx.TiO,04NaSaIt.TiO,10MiMaxx.TiO,14DeHaAl.Re,11MiHuxx.ScO,13SaMaxx.NiO,12SaMaxx.CoH,10SaPaMa.ZnO,11SaMiMa.Re,02KoIsUm.TiH,15LoYuTe.ScH,04KoIsFe.ScH,01TaSeYo.FeH,06KoMaGo.CrH} These studies illustrate the power of the MRCI approach, which has consistently been shown to be a reasonably robust way of investigating low-lying excited states of these systems to usually at least semi-quantitative accuracy when compared against experiment. However, the studies also demonstrate the limitations of even these very high accuracy methods, e.g. correct modelling of higher electronic states and quantitative results. Particularly significant is the large errors found in excitation energies, often in excess of 1000 \cm{} \cite{jt632}. These issues have stimulated new methodological developments in multi-reference methods, e.g. density-matrix renormalisation group theory \cite{05Scxxxx.ai,08MaOnMo.ai,11KuYaxx.ai,11ChShxx.ai,15YaKuMi.ai}, Monte Carlo configuration interaction \cite{14CoMuxx.Re}, and stochastic multi-reference self-consistent field \cite{15ThSuAl.ai}. %These methods, though promising, are still under active development and testing. In particular, dynamic correlation often needs to be added to the method. 

We produce the \abinitio\ data required for a detailed model of the rovibronic spectroscopy of this molecule. The most important {\it ab initio} quantity in the spectroscopic model is the off-diagonal dipole moment, as there is no experimental reference.
 Note that potential energy curves and spin-orbit coupling curves, though important, can be empirically corrected based on experimental VO results.
 The diagonal dipole moment curves for VO will produce the rotational and vibrational transition intensities but these are not as critical for astronomical purposes as the electronic spectra. 
 
In this paper, our methodological discussion focuses on the effect of: 
\begin{itemize}
\item Choice of states used to optimise the complete active space self-consistent field (CASSCF) orbitals  on the icMRCI calculation energies, dipole moment, electronic angular momentum and spin-orbit coupling matrix elements 
\item Finite-field difference vs. expectation value off-diagonal dipole moments %Are the off-diagonal dipole moments calculated using the finite field expression or using expectation values? 
\item Using different vs. identical CASSCF orbitals for off-diagonal
dipole moments and electronic angular momentum matrix elements
\item Inclusion or exclusion of Davidson correction on the final properties 
\end{itemize}

We use icMRCI calculations based on CASSCF orbitals; however, many of our conclusions can be expected to apply, or be extrapolated easily, to other similar multi-reference correlated methods, including the very popular complete active space perturbation theory level 2 (CASPT2), \cite{CASPT2}. 
In this paper, we draw conclusions on best practice in making the above choices. This will be based on theoretical principles, supported by experimental data. 

Further, it is extremely useful to be able to gauge the accuracy of a particular theoretical calculation, i.e. to quantify its uncertainty, even in a non-rigorous approximate way \cite{jt642}. We suggest that a good way of doing this in to evaluate the sensitivity of the given quantity to changed methodology (where the different methodology is slightly inferior on theoretical grounds, but not considerably so). This uncertainty quantification will be useful information to those who use our data, such as experimentalists and astronomers.

\section{Quantitative Study of \emph{Ab Initio} Methodology}

\subsection{Method}

In this section, we will investigate the properties of VO using a few different methodologies in order to assess the best methodology choices, as well as estimate the uncertainty of the results. Our methodologies are chosen to be reasonably modest in cost because we are obtaining curves  rather than single points, and we require results for 13 electronic states. 

In view of time constraints, previous investigations and to focus on current studies, we keep some methodology constant. % a number of methodology considerations. 
Unless otherwise specified, all calculations use aug-cc-pVQZ basis set. We use the
CASSCF \cite{85WeKnxx.ai} implementation within {\sc Molpro}
\cite{12WeKnKn.methods} to find the molecular orbitals that are then
used in {\sc Molpro's} ic-MRCI
\cite{88WeKnxx.ai} program. Note we are using the newly recommended notation \cite{jt632} icMRCI($n$) to indicate $n$ states have been requested in the calculation.

We use a full-valence ($3d$, $4s$/V, $2s$,$2p$/O) active space. Core-correlation effects are quite expensive to include and have been investigated elsewhere \cite{07MiMaxx.VO}. Therefore, we do not consider them in this manuscript. As discussed by Tennyson \cite{jt573}, the
effects of core correlation and relativistic effects often partially cancel each
other in practice. Thus, we will not consider relativistic
corrections in this manuscript in the sense that we do not calculate
Darwin and mass-velocity terms or use the Douglas-Kroll Hamiltonian etc. However,
we do calculate spin-orbit coupling.%, which is also an effect of

State-averaged CASSCF is often used for \emph{ab initio} electronic spectroscopy, commonly without reference to the states used in the state-averaging. However, the large number of electronic states considered for our VO spectroscopic model led us to consider the influence of the way in which the CASSCF orbitals are optimised on the final icMRCI answer. Theoretically, state-specific calculations should give superior results as they are optimised for each individual electronic state (assuming that dynamic correlation introduced post-CASSCF is small). However, the magnitude of this effect is unclear for VO \emph{apriori} and is investigated in this section. %herein for VO. % is unclear apriori what the magnitude  quantitative effect of % because the CASSCF orbitals can 

The standard method for evaluating dipole moments is via an expectation value (XP) of the wavefunction over the dipole moment operator. 
However, it has been found that finite-field difference (FD) expressions can be more accurate for diagonal dipole moments.
The FD expression for a diagonal dipole moment is given by
\begin{equation}
\braket{\Psi|\hat{\mu}|\Psi} = \frac{E (+F) - E(-F)}{2F}
\end{equation}
where $F$ is the strength of the finite electric field, and $E(F)$ is the energy of the molecule in the presence of the electric field.  FD dipoles generally have better convergence
properties than XP dipoles \cite{jt599}, and are usually
closer to experimental values, see Rendell et al.
\cite{87ReBaHu.ai} and references therein.

Almost exclusively, the expectation value expression is used to evaluate off-diagonal dipole moments. However, the finite-field different methodology is also an option. The mathematics were established by Adamson et al. \cite{98AdZaSt.method}, who also discuss a small set of test cases.
Specifically, in symmetried form, we use
\begin{equation}
\label{eq:FF}
\braket{\Psi_1|\hat{\mu}|\Psi_2} =
 (E(\Psi_1)-E(\Psi_2)) \frac{\braket{\Psi_1(+F)|\Psi_2(-F)} - \braket{\Psi_1(-F)|\Psi_2(F)}}{4F}
\end{equation}
where $\Psi(F)$ indicates the wavefunction in the present of the electric field $F$. 
This FD method for off-diagonal dipole moments is more complicated than for diagonal dipole moments and raise other issues which are discussed below. %, and very rarely utilised. 

\subsection{Results}

\subsubsection{Potential Energy Curves}
Several parameters characteristic of the quartet PEC (term energies, $T_e$, experimental vibrational frequencies, 
$T_1$, rotational frequencies, $r_e$ etc.) are known from
experiment \cite{82ChHaMe.VO,82ChTaMe.VO,87MeHuCh.VO,91SuFrLo.VO,94ChHaHu.VO,95AdBaBe.VO,08FiZixx.VO,09HoHaMa.VO}. 
This makes the potential energy curves a useful tool for assessing different levels of theory. The excitation energies and vibrational frequencies are taken from the model Hamiltonian fits for each vibronic band in the original experimental papers. %\alert{When comparing theory to experiment, the dominant error arising from the derived experimental results is probably the way in which the different spin components were averaged. The theoretical results contain no spin differentiation. This effect is likely to be  less than 10 \cm{} for excitation energies and 1 \cm{} for vibrational frequencies, which is below the scale of comparisons we will make below. }

%{1.0}
\begin{sidewaystable}

\caption{\label{tab:ApABPESfit} Quantitative comparison of the quartet state results using different orbitals with icMRCI(n)/aug-cc-pVQZ calculations using ($S$)-CAS orbitals; n=1 for X, A$^\prime$ and D states, 2 for the A state and 3 for the B and C states. Equilibrium bond lengths for the X, A$^\prime$, A, B, C and D states are 1.59, 1.63, 1.64, 1.64, 1.67 and 1.69 \AA{} respectively. Energies and spin-orbit couplings are in \cm{}, dipole moments in units of D and the derivatives are taken in units of \AA$^{-1}$. Experimental values are $r_0$ and $T_0$.} 
\footnotesize
\begin{tabular}{lp{3.5cm}rrrrrrrrrrrrrrrrrrr}%S[round-precision=0]S[round-precision=0]S[round-
\toprule
& $S$ &  \mc{1}{c}{$\tilde{E}_\text{min}$} &  \mc{1}{c}{$\tilde{E}_\text{min}^\text{+Q}$} & \mc{1}{c}{$r_e$} & \mc{1}{c}{$T_1$} & \mc{1}{c}{$T_2$ } &\mc{1}{c}{$r_e^\text{+Q}$} & \mc{1}{c}{$T_1^\text{+Q}$} & \mc{1}{c}{$T_2^\text{+Q}$ } & \mc{1}{c}{$\mu_e^\text{XP}$ } &  \mc{1}{c}{$\mu_e^\text{FD}$ } &  \mc{1}{c}{$\mu_e^\text{FD(+Q)}$} & \mc{1}{c}{$\mu'_e{}^\text{XP}$ } &  \mc{1}{c}{$\mu'_e{}^\text{FD}$ } & \mc{1}{c}{$\mu'_e{}^\text{FD(+Q)}$ } & \mc{1}{c}{SO$_e$ } &  \mc{1}{c}{SO'$_e$ }\\ 
\hline
\X{}	&	(X)	&	0	&	0	&	1.583	&	1028.5	&	2046.5	&	1.586	&	1014.7	&	2018.6	&	-2.50	&	-2.94	&	-3.08	&	-6.13	&	-6.54	&	-6.66	\\									
	&	(X,A$^\prime$,A)	&	918	&	174	&	1.587	&	1025.4	&	2040.7	&	1.587	&	1021.1	&	2031.9	&	-2.41	&	-2.81	&	-2.98	&	-5.21	&	-6.09	&	-6.29	\\									
	&	(X,A$^\prime$,A,B)	&	1053	&	160	&	1.588	&	1017.3	&	2024.6	&	1.587	&	1017.6	&	2025.3	&	-2.58	&	-2.88	&	-3.05	&	-5.07	&	-6.19	&	-6.63	\\									
	&	(X,C)	&	1476	&	319	&	1.587	&	1031.2	&	2052.9	&	1.587	&	1019.6	&	2029.9	&	-2.39	&	-2.90	&	-3.03	&	-6.52	&	-6.12	&	-6.97	\\									
	&	(X,C,D)	&	1382	&	550	&	1.588	&	1021.5	&	2032.6	&	1.588	&	1013.6	&	2017.1	&	-2.01	&	-2.69	&	-2.84	&	-6.14	&	-5.92	&	-6.33	\\									
	&	(X,A$^\prime$,A,B,C)	&	1146	&	191	&	1.588	&	1016.7	&	2023.5	&	1.587	&	1019.2	&	2028.8	&	-2.59	&	-2.87	&	-3.04	&	-4.94	&	-6.13	&	-6.61	\\									
	&	(X,A$^\prime$,A,B,C,D)	&	1555	&	549	&	1.588	&	989.2	&	2006.0	&	1.587	&	997.8	&	1983.0	&	-2.26	&	-2.78	&	-2.99	&	-5.45	&	-8.69	&	-6.69	\\		
	& {MRCI/BP} $^a$  &  &  &1.599 & 983$^*$  & & 1.605 & 980$^*$ & & -2.61 & -3.09  & -3.22   \\	
	& {C-MRCI/BP} $^a$  &  &  &1.589 & 997$^*$ & & 1.592 & 992$^*$ & & -2.43 & -3.00 & -3.23  \\
	& {C-MRCI+DKH2/BP} $^a$  &  &  &1.588 & 1006$^*$ & & 1.591 & 1002$^*$ & & -2.52 & -3.10 & -3.26  \\
	& {(X)-CAS, icMRCI} $^b$ & & & 1.597 & 1006$^\#$ & \\
	& {(X)-CAS, C-MRCI} $^b$ & & & 1.591 & 1020$^\#$ & \\
	&	Exp  \cite{82ChHaMe.VO,82ChTaMe.VO}	&	0	&	0	&	1.5920	&	1001.8	&		&	1.589	&	1001.8	&		&	-3.36	&	-3.36	&	-3.36	\\															
\hline																																								
\Ap{}	&	(A$^\prime$)	&	9854	&	8432	&	1.633	&	941.4	&	1871.0	&	1.629	&	951.0	&	1890.7	&	-2.78	&	-3.28	&	-3.37	&	-5.06	&	-5.51	&	-5.95	&	256.3	&	-47.2	\\%263.987446683675	&	-46.7632	&		
	&	(A$^\prime$,A)	&	9873	&	8443	&	1.633	&	941.8	&	1872.1	&	1.629	&	951.6	&	1892.2	&	-2.77	&	-3.28	&	-3.37	&	-5.05	&	-5.48	&	-5.91	&	254.1	&	-45.9	\\%263.504338833846	&	-45.5181	&		
	&	(A$^\prime$,A,B)	&	10045	&	8515	&	1.632	&	935.2	&	1858.3	&	1.628	&	948.2	&	1885.6	&	-2.90	&	-3.34	&	-3.43	&	-4.90	&	-5.58	&	-6.24	&	237.9	&	-37.0	\\%	&	262.7573166	&	-48.4447	&
	&	(A$^\prime$,A,B,D)	&	10610	&	8909	&	1.636	&	933.3	&	1857.2	&	1.630	&	951.5	&	1894.1	&	-2.59	&	-3.28	&	-3.38	&	-5.10	&	-5.61	&	-6.20	&	240.4	&	-28.6	\\%255.956474858176	&	-42.2303	&		
	&	(X,A$^\prime$)	&	10200	&	8539	&	1.636	&	932.9	&	1843.7	&	1.630	&	946.7	&	1875.8	&	-2.75	&	-3.32	&	-3.40	&	-5.14	&	-5.43	&	-5.89	&	252.1	&	-38.4	\\%262.794950507703	&	-41.569	&		
	&	(X,A$^\prime$,A)	&	10039	&	8497	&	1.634	&	937.7	&	1862.7	&	1.630	&	949.7	&	1887.3	&	-2.75	&	-3.31	&	-3.39	&	-5.11	&	-5.44	&	-5.89	&	241.8	&	-30.6	\\%262.849951183069	&	-43.0605	&		
	&	(X,A$^\prime$,A,B)	&	10076	&	8538	&	1.633	&	934.6	&	1857.5	&	1.629	&	947.4	&	1884.3	&	-2.86	&	-3.35	&	-3.44	&	-4.95	&	-5.60	&	-6.21	&	239.7	&	-33.8	\\%262.610965709621	&	-46.514	&		
	&	(X,A$^\prime$,A,B,C)	&	9821	&	8207	&	1.631	&	942.7	&	1875.0	&	1.627	&	955.5	&	1901.8	&	-2.86	&	-3.34	&	-3.43	&	-4.82	&	-5.68	&	-6.28	&	240.0	&	-37.0	\\%	262.5290659	&	-47.7399	&	
	&	(X,A$^\prime$,A,B,C,D)	&	10339	&	8572	&	1.633	&	853.6	&	1747.4	&	1.627	&	892.0	&	1817.1	&	-2.61	&	-3.31	&	-3.40	&	-4.89	&	-6.07	&	-6.43	&	241.8	&	-30.6	\\%	&	261.8936487	&	-44.9617	
	& {MRCI/B} $^a$  & 7187  & 7084  &1.639 & 913$^*$  & & 1.640 & 912$^*$ & & -3.25 & -3.59  & -3.60   \\
	& {C-MRCI/BP} $^a$  & 8049 & 7753  &1.628 &  & & 1.638 & & & -3.01 &  &   \\
	& {C-MRCI+DKH2/BP} $^a$  & 7956  & 7635  &1.627 &  & & 1.639 &  & &  \\
	& {(A',A,B)-CAS, icMRCI} $^b$ & 7598$^\%$ & & 1.635 & 944$^\#$ & \\
	& {(A',A,B)-CAS, C-MRCI} $^b$ & 7816$^\%$ & & 1.630 & 955$^\#$ & \\
	& {(A',A,B,3$^4\Pi$)-CAS, icMRCI} $^b$ &  7565$^\%$ & & 1.634 & 950$^\#$ & \\
	&	Exp \cite{87MeHuCh.VO}	&	7255	&	7255	&	1.629	&	936.5	&	1865.0	&	1.629	&	936.5	&	1865.0	\\		\botrule
\end{tabular}

$^a$ \emph{Ab initio} results from Miliordos and Mavridis \cite{07MiMaxx.VO}.
$^b$ \emph{Ab initio} results from H\"{u}bner et al. \cite{15HuHoHi.VO}.  

$^*$ $\Delta G_{1/2}$
$^\#$ $\tilde{\nu}_e$
\end{sidewaystable}	

\begin{sidewaystable}
\caption{\label{tab:ApABPESfit2} Continued from Table \ref{tab:ApABPESfit}.}
\footnotesize
\begin{tabular}{lp{3.5cm}rrrrrrrrrrrrrrrrrrr}
\toprule
& $S$ &  \mc{1}{c}{$\tilde{E}_\text{min}$} &  \mc{1}{c}{$\tilde{E}_\text{min}^\text{+Q}$} & \mc{1}{c}{$r_e$} & \mc{1}{c}{$T_1$} & \mc{1}{c}{$T_2$ } &\mc{1}{c}{$r_e^\text{+Q}$} & \mc{1}{c}{$T_1^\text{+Q}$} & \mc{1}{c}{$T_2^\text{+Q}$ } & \mc{1}{c}{$\mu_e^\text{XP}$ } &  \mc{1}{c}{$\mu_e^\text{FD}$ } &  \mc{1}{c}{$\mu_e^\text{FD(+Q)}$} & \mc{1}{c}{$\mu'_e{}^\text{XP}$ } &  \mc{1}{c}{$\mu'_e{}^\text{FD}$ } & \mc{1}{c}{$\mu'_e{}^\text{FD(+Q)}$ } & \mc{1}{c}{SO$_e$ } &  \mc{1}{c}{SO'$_e$ }\\ 																				
\hline																			
\A{}	&	(A)	&	12530	&	10999	&	1.641	&	905.5	&	1797.9	&	1.636	&	920.3	&	1828.5	&	-2.73	&	-3.26	&	-3.35	&	-4.82	&	-5.37	&	-5.89	&	69.0	&	55.1	\\%64.222486717395	&	50.0012	&		
	&	(A$^\prime$,A)	&	12519	&	10988	&	1.641	&	904.8	&	1796.6	&	1.635	&	919.7	&	1827.1	&	-2.74	&	-3.26	&	-3.35	&	-4.82	&	-5.38	&	-5.90	&	69.9	&	55.2	\\%63.981368672799	&	48.764	&		
	&	(A$^\prime$,A,B)	&	12600	&	11014	&	1.641	&	898.3	&	1783.5	&	1.636	&	915.1	&	1818.9	&	-2.91	&	-3.40	&	-3.50	&	-4.77	&	-5.86	&	-6.68	&	55.5	&	72.4	\\%61.121455593422	&	62.0148	&		
	&	(A$^\prime$,A,B,D)	&	13017	&	11224	&	1.646	&	902.3	&	1796.4	&	1.639	&	922.0	&	1837.8	&	-2.60	&	-3.32	&	-3.41	&	-4.92	&	-5.84	&	-6.51	&	60.7	&	66.8	\\%64.542025390125	&	&			
	&	(X,A$^\prime$,A)	&	12665	&	11030	&	1.643	&	900.4	&	1786.8	&	1.636	&	917.2	&	1821.7	&	-2.72	&	-3.28	&	-3.37	&	-4.88	&	-5.34	&	-5.88	&	61.4	&	69.0	\\%63.907242889862	&	49.664	&		
	&	(X,A$^\prime$,A,B)	&	12636	&	11042	&	1.642	&	898.0	&	1783.2	&	1.636	&	914.4	&	1817.7	&	-2.86	&	-3.39	&	-3.49	&	-4.87	&	-5.89	&	-6.64	&	58.1	&	68.2	\\					
	&	(X,A$^\prime$,A,B,C)	&	12454	&	10860	&	1.642	&	902.2	&	1796.7	&	1.638	&	913.6	&	1821.6	&	-2.86	&	-3.38	&	-3.48	&	-4.64	&	-5.85	&	-6.58	&	57.5	&	72.9	\\%61.470878163179	&	60.5446	&		
	&	(X,A$^\prime$,A,B,C,D)	&	12926	&	11195	&	1.637	&	1008.9	&	1944.2	&	1.636	&	960.5	&	1880.1	&	-2.62	&	-3.34	&	-3.42	&	-4.68	&	-6.32	&	-6.72	&	61.4	&	69.0	\\%	&	62.82189598	&	56.8784	
	& {MRCI/B} $^a$  & 9821 & 9614 &1.650 & 866$^*$  & & 1.651 & 866$^*$ & & -3.62 & -3.84  & -3.90   \\	
	& {C-MRCI/BP} $^a$  &  10676& 9677  &1.638 &  & & 1.639 &  & & -3.57   \\
	& {C-MRCI+DKH2/BP} $^a$  & 10669 & 9644  &1.637 &  & & 1.639 &  & & -3.55 & \\
	& {(A',A,B)-CAS, icMRCI} $^b$ & 10138$^\%$ & & 1.644 & 906$^\#$ & \\
	& {(A',A,B)-CAS, C-MRCI} $^b$ & 10259$^\%$ & & 1.639 & 920$^\#$ & \\
	& {(A',A,B,3$^4\Pi$)-CAS, icMRCI} $^b$ & 10106$^\%$& & 1.644 & 910$^\#$ & \\
	&	Exp\cite{82ChTaMe.VO}	&	9499	&	9499	&	1.637	&	884.0	&		&	1.638	&	884.0	&	\\				
\hline																
\B{}	&	(A$^\prime$,A,B)	&	17241	&	13151	&	1.658	&	937.1	&	1879.3	&	1.643	&	952.2	&	1907.7	&	-6.39	&	-6.64	&	-5.99	&	-3.22	&	-7.85	&	-8.74	&	90.6	&	-35.6	\\					
	&	(A$^\prime$,A,B,D)	&	18622	&	13938	&	1.652	&	1051.8	&	2127.7	&	1.634	&	1054.7	&	2132.1	&	-6.24	&	-7.35	&	-6.85	&	-0.84	&	-6.01	&	-7.00	&	89.1	&	-30.2	\\%89.170828160972	&	&			
	&	(X,A$^\prime$,A,B)	&	17630	&	13230	&	1.659	&	951.8	&	1914.0	&	1.641	&	968.7	&	1943.6	&	-6.30	&	-6.85	&	-6.20	&	-3.34	&	-9.24	&	-10.73	&	89.8	&	-30.4	\\					
	&	(X,A$^\prime$,A,B,C)	&	17635	&	13251	&	1.656	&	973.1	&	1966.0	&	1.641	&	984.1	&	1982.7	&	-6.40	&	-6.88	&	-6.23	&	-3.36	&	-9.74	&	-11.40	&	90.8	&	-29.2	\\					
	&	(X,A$^\prime$,A,B,C,D)	&	18790	&	13967	&	1.650	&	1153.8	&	2276.9	&	1.633	&	1126.0	&	2212.8	&	-6.25	&	-7.53	&	-7.05	&	-1.85	&	-17.42	&	-21.17	&	89.6	&	-32.0	\\%	&	92.33834222	&	-29.7812	
	& {MRCI/B} $^a$  & 14743 & 12864  &1.662 & 870$^*$  &  & 1.654 & 894$^*$ & & -6.45 & -6.39  & -5.74   \\	
	& {C-MRCI/BP} $^a$  & 14889 &  13491&1.652 &  & & 1.644 && & -6.75 &  &   \\
	& {C-MRCI+DKH2/BP} $^a$  & 15839  & 14460 &1.650 & 1006$^*$ &  & 1.643 & 1002$^*$ & & -6.82  \\
	& {(A',A,B)-CAS, icMRCI} $^b$ & 14268$^\%$ & & 1.658 & 897$^\#$ & \\
	& {(A',A,B)-CAS, C-MRCI} $^b$ & 15792$^\%$ & & 1.658 & 904$^\#$ & \\
	& {(A',A,B,3$^4\Pi$)-CAS, icMRCI} $^b$ & 14115$^\%$ & & 1.654 & 913$^\#$ & \\
	&	Exp\cite{94ChHaHu.VO, 95AdBaBe.VO}	&	12606	&	12606	&	1.644	&	901.0	&	&	1.644	&	901	\\																								
\hline																																								
\C{}	&	(X,C)	&	19918	&	17222	&	1.665	&	973.4	&	1974.9	&	1.665	&	959.9	&	1947.8	&	-4.47	&	-4.70	&	-4.52	&	-5.82	&	-13.40	&	-13.84	\\									
	&	(X,C,D)	&	20985	&	17635	&	1.677	&	887.6	&	1769.8	&	1.667	&	895.2	&	1790.7	&	-4.50	&	-5.27	&	-4.74	&	-4.00	&	-10.36	&	-12.11	\\				
	& {(X,C,D)-CAS, icMRCI} $^b$ & 17712$^\%$ & & 1.683 & 851$^\#$ & \\
	& {(X,C,D)-CAS, C-MRCI} $^b$ & 19285$^\%$ & & 1.675 & 872$^\#$ & \\		
	& {(X,C,D,3$^4\Sigma^-$,4$^4\Sigma^-$,1$^4\Gamma$)-CAS, icMRCI} $^b$ & 17091$^\%$& & 1.682 & 854$^\#$ & \\
	&	Exp\cite{82ChHaMe.VO}	&	17420	&	17420	&	1.675	&	852.8	&	1699.6	&	1.675	&	852.8	&	1699.6	\\																					
\hline																																								
\D{}	&	(D)	&	18300	&	17310	&	1.679	&	834.2	&	1645.6	&	1.679	&	835.7	&	1649.6	&	-1.31	&	-1.25	&	-1.14	&	-5.48	&	-6.41	&	-6.57	&	&	\\							
	&	(X,C,D)	&	18554	&	17240	&	1.677	&	823.3	&	1603.5	&	1.677	&	833.4	&	1633.2	&	-1.17	&	-1.24	&	-1.14	&	-5.69	&	-5.85	&	-5.96	\\									
	&	(A$^\prime$,A,B,D)	&	20564	&	18033	&	1.699	&	782.6	&	1513.9	&	1.674	&	855.2	&	1689.7	&	-1.61	&	&	&	-6.65	\\											& {(X,C,D)-CAS, icMRCI} $^b$ & 15849$^\%$ & & 1.683 & 850$^\#$ & \\
	& {(X,C,D)-CAS, C-MRCI} $^b$ &  15986$^\%$ & & 1.672 & 863$^\#$ & \\		
	& {(X,C,D,3$^4\Sigma^-$,4$^4\Sigma^-$,1$^4\Gamma$)-CAS, icMRCI} $^b$ &15462$^\%$& & 1.683 & 855$^\#$ & \\		
	&	Exp\cite{87MeHuCh.VO}	&	19148	&	19148	&	1.686	&	836.0	&	&	1.6863	&	836	\\																								
													\botrule
\end{tabular}

$^a$ \emph{Ab initio} results from Miliordos and Mavridis \cite{07MiMaxx.VO}. %CASSCF 
$^b$ \emph{Ab initio} results from H\"{u}bner et al. \cite{15HuHoHi.VO}.  %The basis set is the 
$^*$ $\Delta G_{1/2}$
$^\%$ $T_e$ rather than absolute energy above (X)-CAS, icMRCI(1)/aug-cc-pVQZ minimum. 
\end{sidewaystable}

%\linespread{1.6}

\paragraph{Absolute Minimum Energies}

Quantitative results are given in \Cref{tab:ApABPESfit}.
Energies (usually $E$ or $E_\text{min}$) are given relative to
the energy of the X state using icMRCI(1)/aug-cc-pVQZ with no
relativistic corrections, no core correlation and using orbitals from
(X)-CAS. If the Davidson correction is used, then the energy is given
relative to the X state energy from icMRCI(1)+Q/aug-cc-pVQZ with
(X)-CAS.

 The icMRCI calculation using CASSCF orbitals optimised for fewer states generally gives the lowest icMRCI minimum energy without the Davidson correction. 
There are some subtleties, e.g. the X state energy for (X,C)-CASSCF is higher than using (X,C,D)-CASSCF; however, this is compensated by the rise in C state energy.
%\alert{WHY? WHAT DOES IT MAKE THEM `clearly superior' IF NOT THE LOWEST ENERGY? WHAT IS THE OTHER CRITERIA TO BE }

 With the Davidson correction, the general principle that fewer states in the CASSCF orbitals means lower icMRCI energies still largely holds true,
though the results are more mixed and, of course, the calculations no longer strictly obey the variational principle.
 For the X and A$^\prime$ state, the Davidson correction also dramatically decreases the difference in minimum energy arising from different CASSCF orbitals; without the Davidson correction, the difference in minimum energy is up to 1476 \cm{}, while with the Davidson correction, this decreases to 549 \cm{}. The Davidson correction increases the magnitude of the correlation included. This indicates that post-CASSCF correlation energy is larger when the CASSCF orbitals are less optimised for the particular state.

\paragraph{Excitation Energies}

 By far the biggest error from pure \emph{ ab initio} calculations, is
 the electronic excitation energies, i.e. $T_0$. 
   Without the Davidson
 correction, results in error of more than 5000 \cm{} are observed; the
 \emph{ ab initio} $T_0$ is always larger than the experimental $T_0$.
 The use of the Davidson correction significantly helps, reducing
 errors to generally 1000 - 2000 \cm{} though $T_0$ is still almost always
 overestimated. Clearly,  this is
 far from spectroscopic accuracy. 
 
 It is very important to note that with different CASSCF orbitals for the ground and excited, $T_0$ is no longer uniquely defined. However, they can be inferred by taking differences between the minimum energies. Different choices of CASSCF orbitals do affect the result quite significantly; however, this does not account for our large errors. 
   
    Thus, there is something
 fundamentally missing from the current \emph{ ab initio} treatment of
 VO that causes the higher excited states to be less well described
 than the ground state (resulting in $T_0$ being overestimated). The
 effect of the Davidson correction indicates that inadequate treatment
 of electron correlation may be partially responsible.

\paragraph{Vibrational Frequencies}
The frequency of
the transitions between the $v=0$ and $v=1$, and $v=0$ and $v=2$
vibrational levels are given as $T_1$ and $T_2$ respectively.  Note
that we compare directly with the observed vibrational frequencies, not the
harmonic frequencies; this is done by performing a single-state
nuclear motion calculation for the vibrational energy levels using
{\sc Duo} \cite{jt609}.

The fundamental and first overtone
vibrational frequencies vary by up to 20 \cm{} for the X, A$^\prime$,
A and B states using different CASSCF orbitals. %The difference between the best \emph{ ab initio}
%(taken as the value using state-specific CASSCF orbitals) and experimental frequencies is also less
%than 20 \cm{} for the X and A$^\prime$ states, though it rises to 40 \cm{} for the A state.
Much larger errors, however, are seen for the B, C and D states, up to 100 \cm{}.% These \emph{ ab initio}
%errors in the vibrational spectra are much more severe than for the rotational spectra.

The $T_1$ and $T_2$ change
when the Davidson correction is included, by up to 30 \cm{}. This
change in vibrational energy is fairly systematic within a particular
electronic state across the different choice of CASSCF orbitals.% For
%example, the A$^\prime$ fundamental vibrational frequency increases by
%10-14 \cm{} when the Davidson correction is applied. 
 The non-Davidson
corrected frequencies seem to match experiment better. However, both
are generally higher than the experimentally observed frequencies.

\paragraph{Rotational Frequencies}
Equilibrium bond distances are symbolised by $r_e$. When combined with the reduced mass of a particular isotopologue, these largely determine the rotational frequencies. 
\Cref{tab:ApABPESfit} shows
that  $r_e$ shows variations on the order of 3 m\AA{} with respect to changing CASSCF orbitals. For large $J$ and rovibronic transitions between electronic states with different errors in the rotational constants, this can have a significant influence on the final spectra. However, this is normally not an issue as the rotational constants are usually well defined from experiment. %Further, for VO, we actually don't use the equilibrium bond distances from theory for any of the main absorption bands as we fit to experiment. 
%with variation usually less than 0.003 \AA{}
%(corresponding to a change in the rotational constant of 0.002 \cm{}),
%and always less than 0.03 \AA{} (rotational constant error less than 0.02 \cm{}).
%Larger errors occur for the higher lying electronic states. 
%Errors in the transition frequencies above 1 \cm{}
%thus usually occur only for $J>$250 and even in the worst case for $J>$25. Thus, the \emph{ ab initio} error
%in the rotational constant is small when considering vibronic and rovibrational transitions in the visible and infrared.

\subsubsection{Diagonal Dipole Moments}

\begin{figure}
\includegraphics[width=\textwidth]{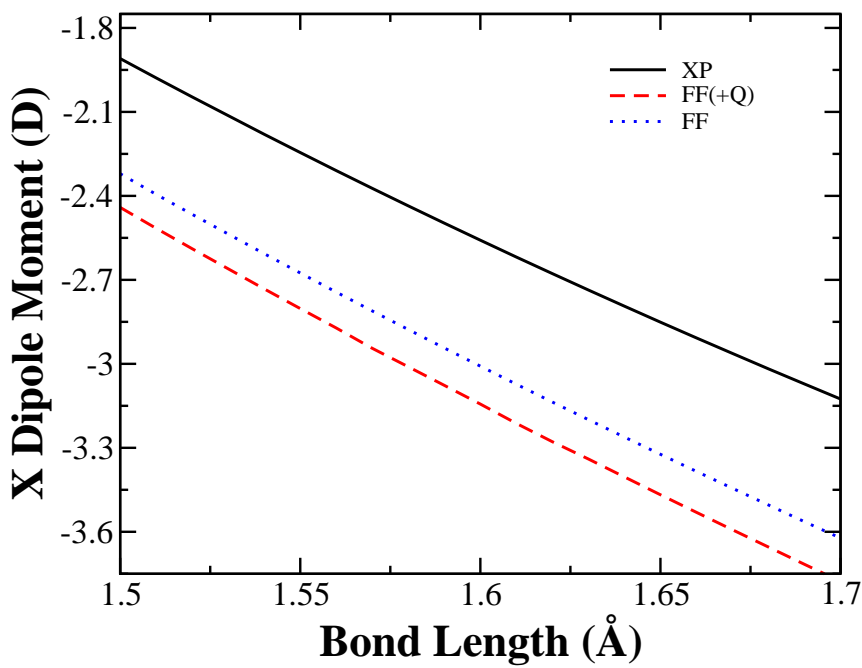}
\caption{\label{fig:X_DM} VO \X{} state diagonal dipole moments, computed using icMRCI(1)/aug-cc-pVQZ with (X)-CAS.}
\end{figure}

\Cref{fig:X_DM} shows that for the \X{} state, the FD diagonal dipole
moment using (X)-CAS orbitals,  both
with the Davidson correction (-3.08 D at equilibrium) and without (-2.94 D at equilibrium), is
significantly more negative than the XP one (-2.50 D at equilibrium).
The experimental value is -3.355 $\pm$
0.014 D \cite{91SuFrLo.VO}; the FD results are closer to this value. Note that this comparison is not strictly fair, as the experimental dipole moment is a vibrationally averaged quantity which we compare to the static equilibrium dipole, $\mu_e$; however, this should be reliable to within the errors discussed here. Further, spin-orbit coupling means higher excited states (and thus their dipole moments)  could contribute to the lowest rovibronic state. However, since the X state is well separated from all other electronic states, this effect is very small. 

\begin{figure}
\includegraphics[width=\textwidth]{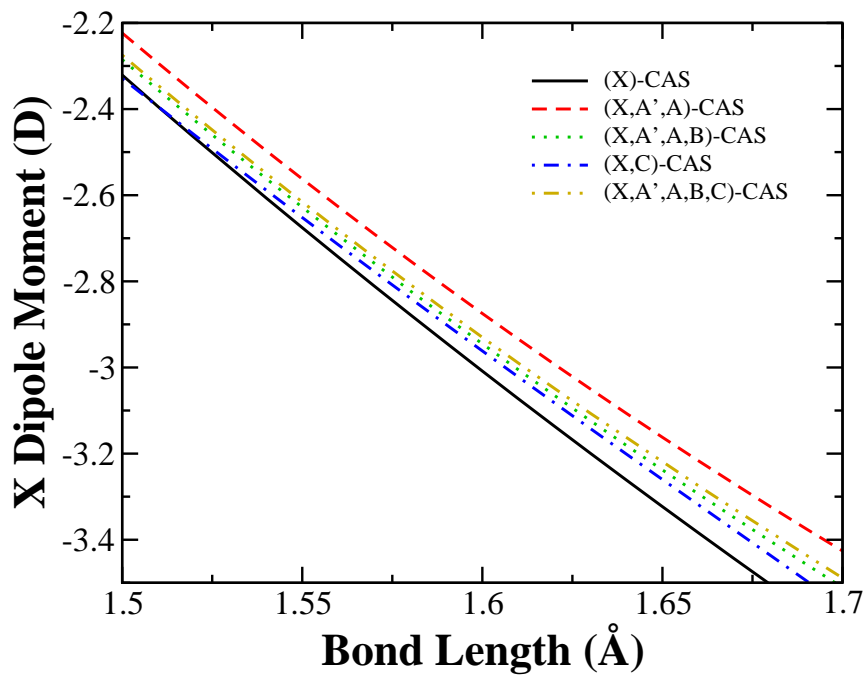}
\includegraphics[width=\textwidth]{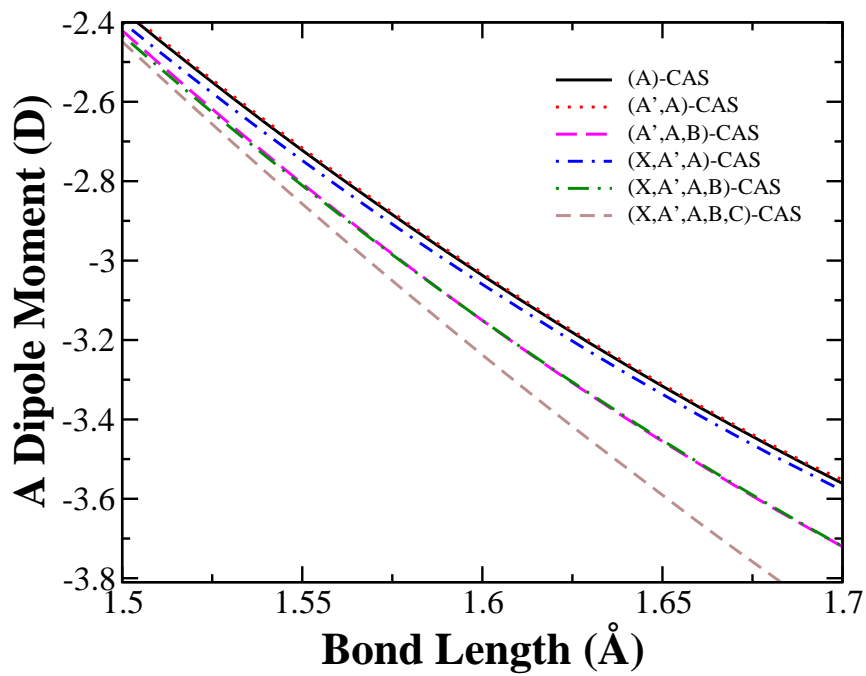}
\caption{\label{fig:X_DM_FF_noDC} VO \X{} (top)  and \A{} (bottom) state diagonal dipole moment, evaluated with FD methodology, using icMRCI(1)/aug-cc-pVQZ  and icMRCI(2)/aug-cc-pVQZ energies respectively.}
\end{figure}

\Cref{fig:X_DM_FF_noDC} shows the difference in the
DMC for the \X{} state when evaluated using different
CASSCF orbitals, with FD methodology and icMRCI(1)/aug-cc-pVQZ energies.
The curves all have the same general shape, but they are off-set
vertically and there is some deviation from parallel curves. 
\Cref{fig:X_DM_FF_noDC} also shows the difference in the DMC for the \A{} state
when evaluated using different CASSCF orbitals, with FD methodology and icMRCI(2)/aug-cc-pVQZ energies.
Apart from the CASSCF orbitals averaged over more than 5 states, the results separate clearly into two curves,
based on whether the B state is included in the CASSCF orbital optimisation. From this result,
it is apparent that the A and B state have significantly different charge distributions.
%As mentioned previously, it is unclear which wavefunction is better; instead this should be seen as indicative of the error in the dipole moment.

\Cref{tab:ApABPESfit} compares the quartet states diagonal dipole moments calculated using different orbitals. %between different diagonal dipole moments for all quartet states.
Generally different orbitals give dipole moments that vary by up to 15 \%.
The derivatives have larger variation, up to approximately 30 \%.
The finite field dipole moments and derivatives are generally about 0.5 D and 0.5 D/\AA{} respectively
lower than the expectation value dipole moments and derivatives.

\begin{table*}
\caption{\label{tab:offdiagDM} Off-diagonal dipole moment matrix elements evaluated using icMRCI($n$)/aug-cc-pVQZ using (Bra $S$)-CAS for bra wavefunction and (Ket $S$)-CAS orbitals for the ket wavefunction; $n$ is 1 for X, 2 for A and C, 3 for B. The equilibrium bond length is taken as 1.59 \AA{} in all cases, for ease of comparison. Dipole moments in D and derivatives taken in units of D/\AA$^{-1}$.}
\footnotesize
\begin{tabular}{lllrrrrrr}
\toprule
& Bra $S$  & Ket $S$& \mc{1}{c}{${\mu_e}$} & \mc{1}{c}{${\mu_e}'$} \\
\hline
X-A	&	(X)	&	(A)	&	0.48	&	-0.60	\\%&	1.32	&	0.51	\\				
X-A	&	(X)	&	(A$^\prime$,A)	&	0.47	&	-0.59	\\%&	1.29	&	0.49	\\				
X-A	&	(X)	&	(A$^\prime$,A,B)	&	0.65	&	-0.04	\\%&	1.29	&	0.67	\\				
X-A	&	(X)	&	(X,A)	&	0.44	&	-0.50	\\%&	1.33	&	0.48	\\				
X-A	&	(X)	&	(X,A$^\prime$,A)	&	0.46	&	-0.55	\\%&	1.33	&	0.49	\\				
X-A	&	(X)	&	(X,A$^\prime$,A,B)	&	0.54	&	-0.61	\\%&	1.29	&	0.53	\\				
X-A	&	(X,A)	&	(A)	&	0.47	&	-0.58	\\%&	1.36	&	0.54	\\				
X-A	&	(X,A)	&	(A$^\prime$,A)	&	0.46	&	-0.57	\\%&	1.32	&	0.51	\\				
X-A	&	(X,A)	&	(A$^\prime$,A,B)	&	0.53	&	-0.59	\\%&	1.32	&	0.55	\\				
X-A	&	(X,A)	&	(X,A)	&	0.43	&	-0.48	\\%&	1.37	&	0.51	\\
X-A	&	(X,A)	&	(X,A$^\prime$,A)	&	0.45	&	-0.53	\\%&	1.36	&	0.52	\\				
X-A	&	(X,A)	&	(X,A$^\prime$,A,B)	&	0.52	&	-0.59	\\%&	1.36	&	0.57	\\				
X-A	&	(X,A$^\prime$,A)	&	(A)	&	0.47	&	-0.57	\\%&	1.36	&	0.54	\\				
X-A	&	(X,A$^\prime$,A)	&	(A$^\prime$,A)	&	0.45	&	-0.56	\\%&	1.33	&	0.52	\\				
X-A	&	(X,A$^\prime$,A)	&	(A$^\prime$,A,B)	&	0.53	&	-0.58	\\%&	1.32	&	0.55	\\				
X-A	&	(X,A$^\prime$,A)	&	(X,A)	&	0.42	&	-0.47\\%	&	1.37	&	0.51	\\				
X-A	&	(X,A$^\prime$,A)	&	(X,A$^\prime$,A)	&	0.44	&	-0.52	\\%&	1.37	&	0.52	\\				
X-A	&	(X,A$^\prime$,A)	&	(X,A$^\prime$,A,B)	&	0.51	&	-0.58	\\%&	1.36	&	0.57	\\				
X-A	&	(X,A$^\prime$,A)	&	(X,A$^\prime$,A)	&	0.49	&	-0.54	\\%&	1.36	&	0.55		\\
X-A	&	(X,A$^\prime$,A,B)	&	(A$^\prime$,A,B)	&	0.65	&	-0.05	\\%&	1.33	&	0.68	\\				
X-A	&	(X,A$^\prime$,A,B)	&	(X,A$^\prime$,A,B)	&	0.70	&	0.00	\\%&	1.36	&	0.65		\\
X-A	&	(X,A$^\prime$,A,B,C)	&	(X,A$^\prime$,A,B,C)	&	0.64	&	-0.16	\\%&	1.37	&	0.66		\\
X-A	&	(X,A$^\prime$,A,B,C,D)	&	(X,A$^\prime$,A,B,C,D)	&	0.73	&	-0.21	\\%&	1.36	&	0.73		\\
\hline																	
X-B	&	(X)	&	(A$^\prime$,A,B)	&	1.51	&	-2.86\\%	&	0.01	&	0.25	\\				
X-B	&	(X)	&	(X,A$^\prime$,A,B)	&	1.52	&	-2.82	\\%&	0.05	&	0.42	\\				
X-B	&	(X,A)	&	(A$^\prime$,A,B)	&	1.53	&	-2.83	\\%&	0.01	&	0.11	\\				
X-B	&	(X,A)	&	(X,A$^\prime$,A,B)	&	1.53	&	-2.78	\\%&	0.02	&	0.27	\\				
X-B	&	(X,A$^\prime$,A,B)	&	(A$^\prime$,A,B)	&	1.51	&	-2.83\\%	&	0.02	&	0.13	\\				
X-B	&	(X,A$^\prime$,A,B)	&	(X,A$^\prime$,A,B)	&	1.67	&	-3.12\\%	&	0.01	&	0.17		\\
X-B	&	(X,A$^\prime$,A,B,C)	&	(X,A$^\prime$,A,B,C)	&	1.55	&	-2.86	\\%&	0.05	&	0.33	\\
X-B	&	(X,A$^\prime$,A,B,C,D)	&	(X,A$^\prime$,A,B,C,D)	&	1.67	&	-3.55\\%	&	0.01	&	0.13		\\
\hline																	
X-C	&	(X)	&	(X,C)	&	2.92	&	-3.76	&	\\							
X-C	&	(X)	&	(X,C,D)	&	2.88	&	-2.93	&	\\							
X-C	&	(X,C)	&	(X,C)	&	3.03	&	-3.11	\\								
X-C	&	(X,C)	&	(X,C,D)	&	2.99	&	-2.20	\\								
X-C	&	(X,C,D)	&	(X,C,D)	&	3.08	&	-2.70	\\								
\hline																	
A-B	&	(A)	&	(A$^\prime$,A,B)	&	0.17	&	-1.23	&	\\							
A-B	&	(A)	&	(A$^\prime$,A,B,D)	&	0.10	&	-1.87	&	\\							
A-B	&	(X,A)	&	(A$^\prime$,A,B)	&	0.19	&	-1.24	&	\\							
A-B	&	(X,A)	&	(X,A$^\prime$,A,B)	&	0.22	&	-1.01	&	\\							
A-B	&	(A$^\prime$,A,B)	&	(A$^\prime$,A,B)	&	0.38	&	-0.59	&			\\		
A-B	&	(A$^\prime$,A,B,D)	&	(A$^\prime$,A,B,D)	&	0.39	&	-0.64	&			\\		
A-B	&	(X,A$^\prime$,A,B)	&	(X,A$^\prime$,A,B)	&	0.38	&	-0.48	&			\\		
A-B	&	(X,A$^\prime$,A,B,C)	&	(X,A$^\prime$,A,B,C)	&	0.37	&	-0.45	&	&		\\		
A-B	&	(X,A$^\prime$,A,B,C,D)	&	(X,A$^\prime$,A,B,C,D)	&	0.38	&	-0.49	&	&		\\		
\botrule																	
\end{tabular}
\end{table*}

\begin{figure}
\includegraphics[width=\textwidth]{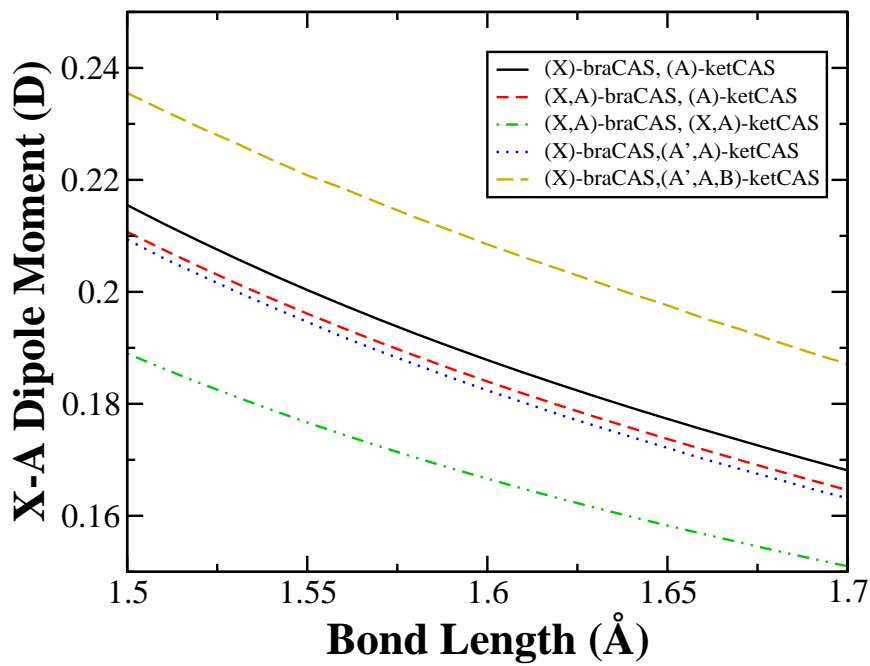}
\caption{\label{fig:offdiagDMCXApAB} X-A off-diagonal dipole moment, bra wavefunction (of \X{} state) from icMRCI(1)/aug-cc-pVQZ with braCAS orbitals, ket wavefunction (of \A{} state) from icMRCI(2)/aug-cc-pVQZ with ketCAS orbitals.}
\end{figure}

\subsubsection{Off-Diagonal Dipole moments}

\paragraph{Finite-field}

\begin{table*}
\caption{\label{tab:offdiagFF} Comparison of off-diagonal dipole moment of VO at a bond distance of 1.59 \AA{}, using different methodologies, specified in text. Energies are given in \cm{}, dipole moments in D. 
}
\begin{tabular}{llllrrrrrrrrrrrrr}%S[round-precision=0]S[round-precision=3]S[round-precision=1]S[round-precision=3]S[round-precision=2]S[round-precision=2]S[round-precision=1]S[round-precision=2]S[round-precision=1]}
%\hline\hline
\toprule
& Symmetry & Method & & \mc{1}{c}{$\Delta E$} &  \mc{1}{c}{${\mu_e}$}  & \mc{1}{c}{Time (min)} \\%& \mc{1}{c}{$\abs{SO_e^*}$ } & \mc{1}{c}{$\abs{SO_e^*}'$} \\
\hline
X-A  & C$_{\infty v}$&  CASSCF(1,1,1,0) & XP & & 0.267 & 3.6  \\
& C$_1$ &  CASSCF(5) & XP & & 0.267 & 6.1 \\
& C$_1$ &   CASSCF(5) &FD & 13879.4 & 0.359  & 9.0 \\%341.90 \\
&  C$_1$ & CASSCF(5)+Exp.E &FD & 9498.9 & 0.245  & " \\
& C$_{2v}$ & icMRCI(1,2,2,0) & XP & & 0.492 & 16.0 \\
& C$_1$ & icMRCI(5) & XP & & 0.507 & 421.9 \\
& C$_1$& icMRCI(5)& FD & 12154.0 & 0.546 & 1321.4\\%3938.48 \\
& C$_1$& icMRCI(5)+Exp.E & FD & 9499 & 0.427  & "\\%3938.48 \\
& C$_1$& icMRCI(5)+Q & FD & 10931.55 & 0.491  & "\\%3938.48 \\
\hline
X-C & C$_{\infty v}$ & CASSCF(2) &  XP & & 2.779 & 2.8 \\%117.38 \\
& C$_{\infty v}$&  CASSCF(2) &  FD& 24847.6&3.710 & 3.3 \\%146.08 \\
& C$_{\infty v}$ &  CASSCF(2)+Exp.E &  FD& 17420.1&2.6& " \\
 & C$_{2 v}$ &  icMRCI(2) &  XP &  & 3.309 & 10.8 \\%8670.87   \\
 & C$_{2 v}$ &  icMRCI(2) &  FD &   19397.1 &   3.712  & 27.0 \\%25683.49\\
 & C$_{2 v}$ &  icMRCI(2)+Exp.E &  FD & 17420.1  &  3.334  &"\\
 & C$_{2 v}$ &  icMRCI(2)+Q &  FD &  17739.8 & 3.395   &"  \\
\hline\hline
\end{tabular}
\end{table*}

We report calculations of the off-diagonal FD dipole moments compared to XP for VO in \Cref{tab:offdiagFF}, giving details of both numerical results and timings. 

\subsubsection*{Parallel Transitions} 
FD calculations for parallel transitions, typified by the X-C transition, are practical in terms of calculation time. The use of FD rather than XP triples the icMRCI calculation time. 

However, it is unclear what is the best choice for the energies of
the zero-field wavefunctions; with or without Davidson correction, with the same orbitals or different orbitals for each state,
using experimental values for energies, etc. This has a substantial effect of the magnitude of the dipole moment as it is a pure multiplicative factor. Though one of these FD results may in fact be  more accurate than the XP result, it is currently unclear which result this would be. 

\begin{figure}
\includegraphics[width=\textwidth]{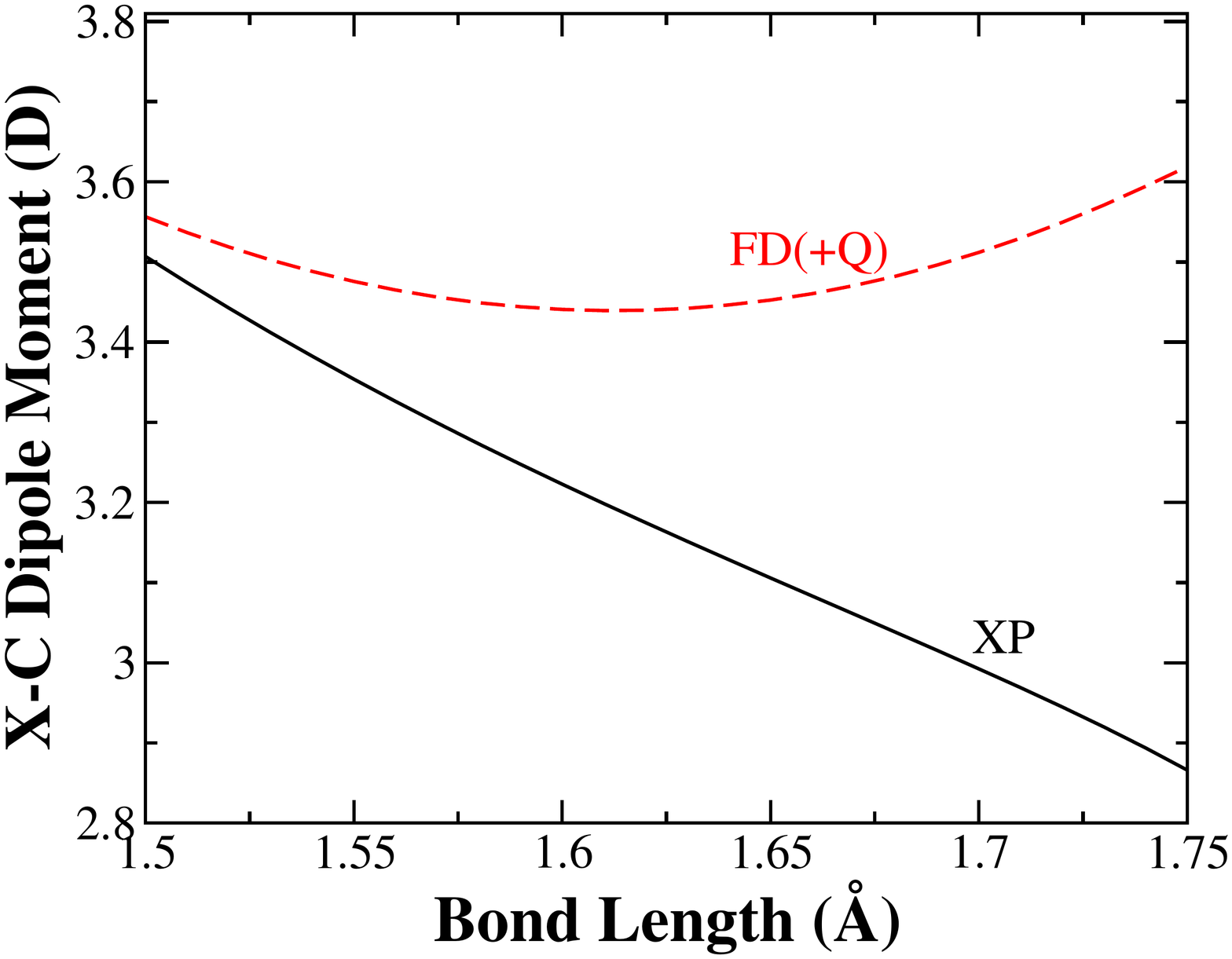}
\caption{\label{fig:offdiagFF} \X-\C{} off-diagonal dipole moment, using icMRCI(2)/aug-cc-pVQZ with (X,C)-CAS for bra and ket wavefunction. The solid line is calculated using expectation value method, the dotted line is finite field method using energies with Davidson correction.}
\end{figure}

 One further cause for concern for the off-diagonal FD dipole moments is consideration of their behaviour as a function of bond length,
shown in \Cref{fig:offdiagFF}. The XP curve goes smoothly towards zero at long bond lengths (as expected for this transition). However, some FD results seem to increase with increasing bond length. The dipole moment may eventually turn over at long bond lengths,
but this behavior does not inspire confidence in the FD results.

\subsubsection*{Perpendicular Transitions} 

For perpendicular transitions, the finite field needs to be applied perpendicular to the molecular axis which breaks the
linear-molecule symmetry. This means  C$_1$ symmetry must be used: all electronic states  of a given spin and both components
of $\Pi$, $\Delta$, $\Phi$ etc. states have to be included in the calculation. For
example, for the X-A dipole, we need 5 states.
\Cref{tab:offdiagFF} presents some CASSCF/aug-cc-pVDZ and icMRCI/cc-pVDZ results. For the X-A dipole moment, the reduction in symmetry increases the calculation time more than 25 fold for icMRCI(5) in C$_{1}$ vs icMRCI(1,2,2,0) in C$_{2v}$ symmetry for the XP method, and even more for the FD method (calculation times around 80 times longer), due to the difficulty of converging the larger number of states and increased number of reference states in the icMRCI.

\paragraph{Expectation value}

Given the difficulties experienced with the finite-field off-diagonal dipole moments and the importance of this parameter to the final spectroscopic model, we investigate a range of different CAS orbitals here. 
\Cref{fig:offdiagDMCXApAB} shows the X-A off-diagonal dipole moment
curves for five choices of bra and ket CAS orbitals calculated using the expectation value methodology.
\Cref{tab:offdiagDM} provides quantitative comparison metrics for the
X-A, X-B, X-C and A-B off-diagonal expectation value dipole moments. %Of the three main bands, the magnitude of the X-C transition moment is largest, while the magnitude of the X-A transition moment is smallest. %The electronic angular momentum coupling between the X and A

In most cases, different orbital choices give qualitatively similar results for both the off-diagonal dipole moment and electronic angular momentum matrix elements. A variation of 10 \% between the equilibrium off-diagonal dipole moments due to different CASSCF choices is usual. %The X-A electronic angular momentum matrix element at equilibrium is between 1.287 and 1.373 $\hbar$, showing variation of only 6 \%. 
%The variability of the X-B element between 0.008 and 0.051 $\hbar$ is small in absolute terms but large in relative terms. 
The derivatives are more variable with deviations of more than 40 \% seen for different CASSCF choices in the X-C dipole moment derivative ( -2.20 to -3.76 D/\AA{}).% and the X-A electronic angular momentum derivative (0.48 to 0.73 $\hbar$/\AA{}).

However, there are some cases where there is a much larger, qualitatively significant, effect.
In particular, it is clear from the magnitude of the X-A and A-B dipole moments and their
derivatives that the inclusion of the B state in the CASSCF orbital optimisation has
a large effect on the electron distribution in the final icMRCI A wavefunction.
For example, the X-A dipole moment increases by more than 25 \%  and the A-B dipole moment doubles. The absolute dipole moment increases by approximately 0.2 D in each case.
An even larger effect is seen in the derivatives of the X-A dipole moment which is reduced
dramatically from about 0.5 D/\AA{}  to near zero when the B state is included.
Again we see that the B state must have a fundamentally different character to the A state.

\begin{table}
\caption{\label{tab:offdiagDM} 
Off-diagonal spin-orbit matrix element calculations using CASSCF/aug-cc-pVDZ using ($S$)-CAS. The equilibrium bond length is taken as 1.59 \AA{} in all cases, for ease of comparison. Spin-orbit couplings are given in \cm{} and derivatives taken in units of \cm{}/\AA.}
\footnotesize
\begin{tabular}{lllrrrrrr}
\toprule
& $S$  & \mc{1}{c}{${SO_e}$ } & \mc{1}{c}{${SO_e}'$} \\
\hline									
X-A	&	(X,A)	&		68.9	&	10.5	\\
X-A	&	(X,A$^\prime$,A)	&	69.8	&	11.1	\\
X-A	&	(X,A$^\prime$,A,B)	&	66	&	15.4	\\
X-A	&	(X,A$^\prime$,A,B,C)	&	66.5	&	15.4	\\
X-A	&	(X,A$^\prime$,A,B,C,D)	&	65.3	&	23.7	\\
\hline									
X-B	&	(X,A$^\prime$,A,B)	&	7.9	&	45.4	\\
X-B	&	(X,A$^\prime$,A,B,C)	&	7.5	&	46.4	\\
X-B	&	(X,A$^\prime$,A,B,C,D)	&	5.3	&	43.7	\\
\hline									
A-B	&	(A$^\prime$,A,B)	&	2.7	&	-2.8	\\
A-B	&	(A$^\prime$,A,B,D)	&	3.2	&	-1.9	\\
A-B	&	(X,A$^\prime$,A,B)	&	2.2	&	-2.7	\\
A-B	&	(X,A$^\prime$,A,B,C)	&	2.2	&	-3	\\
A-B	&	(X,A$^\prime$,A,B,C,D)	&	2.6	&	-3	\\
\botrule																	
\end{tabular}
\end{table}

\subsubsection{Spin-orbit coupling parameters}
Previous benchmarking on ScH  \cite{jt599} suggested that CASSCF/aug-cc-pVDZ calculations were sufficiently accurate for the construction of line lists for TM diatomics.%, especially considering the other sources of error and the fact that empirical corrections are usually applied to refine the model to experimental results. Thus here we will consider this methodology.
We follow this method and basis set, and focus on the influence of the electronic states included in the CASSCF calculation. Note that for spin-orbit coupling elements, the bra and ket wavefunctions must use the same CASSCF orbitals.

\Cref{tab:ApABPESfit} compares the diagonal spin-orbit coupling matrix elements for all non-singlet quartet states.
\Cref{tab:offdiagDM} provides comparison metrics of some off-diagonal quartet spin-orbit couplings matrix elements at equilibrium.
Both these comparisons demonstrate that different choices of states in the CASSCF affects the matrix element by
roughly 3-5 cm$^{-1}$ in absolute magnitude. This is similar to that observed for the variation between different basis sets and icMRCI versus CASSCF for ScH  \cite{jt599}.
%; this provides an estimate of the error in this parameter.

\subsection{Discussion}

\subsubsection{Davidson correction}
The Davidson correction\cite{74LaDaxx.ai,86BaLaTa.ai} (conventionally indicated by +Q) provides an estimate of the triples and quadruples contribution to the correlation energy, improving the size-consistency of the calculation. 
Previous icMRCI studies on related TM oxides show that the Davidson
correction influences the excitation energy significantly, up to 2000
\cm\ \cite{07MiMaxx.VO,10MiMaxx.TiO,11SaMiMa.Re}; often (but not
always) this is a decrease in energy that improves agreement with
experiment.  These studies show changes in bond length, harmonic
vibrational frequency and diagonal dipole moment (computed
using finite fields) are fairly small: about 0.005 \AA, 10 \cm\ and
0.1 D respectively. However, it is not clear whether
the Davidson correction improves results comparing to experiment. Our results provide additional quantitative data on this effect, illustrating the generality of the previous results. 

\subsubsection{Choice of Orbitals for Multi-reference calculations}

The results in this section demonstrate that the way in which the CASSCF orbitals are optimised, specifically what electronic states are considered, has a profound effect on the final icMRCI calculation. 
 A particularly striking example seen in VO is that inclusion of the B state in
optimising CASSCF orbitals for the icMRCI calculations of the X, A$^\prime$ and A states significantly affects the DMC. 

On a semi-quantitative level, the property curves obtained using
different orbitals provides an
estimate of the uncertainty in the prediction of the property. The `best'
answer is taken as the result using the CASSCF
orbitals optimised for the smallest number of electronic states.

From a theoretical perspective, generally optimising for as few states as possible is preferred as the CASSCF orbitals more closely match the wavefunction of the electronic state under consideration. This will usually give lower icMRCI energies and, presumably (but not certainly), better properties \cite{88BaLaxx.AlH}.
However, there are at least three cases where we believe that extra electronic states must be included.
\begin{enumerate}
\item For higher states of a particular spin and symmetry, the lower energy states should be included.% without including the
%  lowest energy states in the SA-CASSCF and icMRCI calculations.  That is, to
 % find results for the \B{} state, one must include the \A{} state in
  %the CASSCF and icMRCI calculations also. This is because otherwise the
  %\B{} state would fall in energy in the calculation to become the
  %\A{} state.
\item For icMRCI calculations and low-symmetry C$_{2v}$ SA-CASSCF calculations, all states of lower energy with the same $C_{2v}$ symmetry must be included.%; e.g. the \Ap{} state must be included in C$_{2v}$ calculations of the \B{} state.
\item Sometimes the order of electronic states is not correctly reproduced by calculations; therefore states of slightly higher energy can be useful or necessary. %For example, calculations often give incorrect ordering of the \C{} and \D{} states.
\end{enumerate}
Note that both components of the $\Lambda\ne 0 $ states should be included in C$_{2v}$ SA-CASSCF calculations, for example, $A_1$ and $A_2$ in case of the  $\Delta$ symmetry.

Ideally, there would be no reason to include extra states. However, due to the high nonlinearity of the
problem, CASSCF has substantial convergence problems
\cite{10SlMaxx.ai} and discontinuities in the PEC and/or DMC are
common \cite{88BaLaxx.AlH,90MeLeMa.ai,94ZaMaxx.ai,97GuMaMa.ai,03AnCaCi.ai}.
Therefore, extra states in the
CASSCF orbital optimisation also can be useful because
\begin{enumerate}
  \setcounter{enumi}{3}
\item Increasing the SA-CASSCF space can provide better convergence
  \cite{98ScGoxx.ai}. For example it can help avoid root flipping
  \cite{72DoHixx.LiH}.
 \item SA-CASSCF helps recovering the degeneracy at the dissociation limit \cite{97SuAsKo.Sc2}.
% \item The SA-CASSCF orbitals are expected to better represent the mixing of the two configurations for all distances of $r$ and provide  qualitatively correct potential curves \cite{97SuAsKo.Sc2}.
%\item The inclusion of upper state of the same symmetry may improve the representation of the lower states (see below).
\end{enumerate}
%,94ZaMaxx.ScH,90SaLeMa.VO,94ZaMaxx.ScH,13ShLexx.VO}.

Finally, there are some pragmatic reasons why SA-CASSCF may be preferred:
\begin{enumerate}
  \setcounter{enumi}{5}
\item Running calculations based on a single SA-CASSCF orbitals for all properties of interest is faster in terms of human time set-up costs.
\item To link icMRCI calculations based on SS- or MS-CASSCF orbitals, it is useful (and sometimes essential) to run SA calculations to find the correct phases and signs of wavefunctions and matrix elements. %State-averaged calculations ar \item State averaging is useful for computing couplings  among electronic states.
\end{enumerate}

Even with SS- and MS-CASSCF orbitals, icMRCI calculations do not give satisfactory answers for many properties.
Of particular concern is the very large errors in the electronic excitation energies, sometimes by up to 5000 \cm{}. 
Even with the Davidson correction, non-systematic over-estimates of about 1000 \cm{} are the norm. 
Furthermore, even the qualitative ordering of states is often wrong, e.g. the \C{} and \D{} states in VO are often switched. 
%and the nature and ordering \Df, \Dg{} and \Dh{} are so variable that we rely primarily on experimental evidence of the \Df{} and \Dg{} states, and ignore the \Dh{} state entirely, in producing our final spectroscopic model. 
Aside from the obvious effect that these incorrect excitation energies have on the
band origins, there are also more subtle effects. In particular, perturbations between observed (``bright'') and hidden (``dark'') states are often seen as resonance interactions between electronic states via, for example,
off-diagonal spin-orbit interactions. Characterizing these resonances for example by assigning electronic and vibrational states to the dark state is often difficult experimentally. Unfortunately it would appear that at present \emph{ ab initio} methods remain unable to help with this for molecules like VO. In contrast, for main-group molecules such as C$_2$ ab-initio methodologies are greatly facilitating the identification of new spectral bands, e.g. \cite{15KrBaTr.C2}. 
%\red{Not sure of the hence leading this paranthetic phrase. Can it be deleted?} (hence why CASSCF/aug-cc-pVDZ is deemed sufficient).

\subsubsection{Dipole Moments}%Choice of Orbitals for Multi-reference calculations}

Worse than the errors in excitation energy, which at least can be generally quantified, are the unknown errors in the dipole moment curves. For VO, only the ground experimental value is known. Lifetimes have been measured for certain states \cite{97KaLiLu.VO} which serve as a useful verification of \emph{ ab initio} results;\cite{jt624} more of these measurements would be helpful.

There are various theoretical arguments \cite{81DiRoSa.ai,92ErMaPe.ai,02Lixxxx.ai,jt475} to expect that dipole moments computed using finite field differences (FD)
are superior to those obtained using expectation values (XP).
However, Erzernor et al.
\cite{92ErMaPe.ai} show that there are certain requirements that
must be fulfilled for the FD to be superior; in particular, the optimal energy must be
obtainable with only second-order corrections to the orbitals and CI
coefficients etc.  For diagonal dipole moments, these conditions are
generally fulfilled and theoretical argument about the superiority of
the FD dipoles agrees with what is found in practice in \emph{ ab
  initio} calculations. This is also true for the ground
state of VO \cite{07MiMaxx.VO}. However, our results show that it is probably not so straightforward for off-diagonal dipole moments. 

Our results for the diagonal dipole moments confirm the advantage of using the finite-field differences rather than expectation values; automating this process within quantum chemistry packages with new keywords would facilitate the use of this methodology.
%These results suggests that quantum chemistry programs should include automated routines for calculating finite field diagonal dipole moments and these should always be used in preference to expectation values.
 Note that CFOUR \cite{09StGaHa.ai} has such routines for coupled-cluster related methodologies.

Given the importance of the off-diagonal dipole moments to the quality of the final line list, it is important that we maximise the accuracy of the expectation value results. Finite-field methodology seems a logical choice; however, the success of the finite-field equation for diagonal dipole moments is not matched for off-diagonal dipole moments.  There are two specific issues. First, should the same electronic structure method be used to calculate the energies and wavefunctions of the two states? Transition energies are notoriously badly predicted by theory for transition metal diatomics; errors of 1000 \cm{} or more are regularly encountered \cite{jt632}. Without the Davidson correction (which improves the energy but does not correct the wavefunction), for VO we found errors in excess of 5000 \cm{}. Another option is to use the experimental excitation energy. 
Second, the application of a finite electric field for a perpendicular transition in a diatomic molecule (e.g. between a sigma and pi electronic state) reduces the symmetry of the calculation. This means that a larger number of states need to be included in the calculation. This significantly increases calculation time and decreases the accuracy of the result.  

 Given these issues with finite-field off-diagonal dipole moments, we must use the expectation value method. However, we argue that a significant improvement can be obtained by representing the icMRCI wavefunction of both the bra and ket wavefunctions using different, optimised CASSCF orbitals for each electronic state. Without a higher-accuracy 'near-exact' result, we cannot demonstrate that the result is always improved. However, we can and do demonstrate that the variation of the off-diagonal dipole moments obtained using different CASSCF orbitals routinely vary by approximately 10\%, with more extreme errors of up to 100\% found when the CASSCF orbitals are trying to represent two states with very different charge distributions (e.g. when considering the \A{} and \B{} state in VO).

\section{Final Calculations}% Results}
\subsection{Method}
For the final calculation for the spectroscopic model of VO, the electronic structure calculations were generally 
icMRCI/aug-cc-pVQZ, except for the a-g, c-g and d-g off-diagonal dipole moments where CASSCF calculations were used. As far as possible, we use SS-CASSCF calculations to obtain the
orbitals for icMRCI.  
The \B{} state wavefunction used (A',A,B)-CAS, the \Dg{} state wavefunction used (e,f,g)-CAS and the \C{} state used (X,C)-CAS.  %For all other electronic states, no additional states are required in the CASSCF. 
icMRCI(1) calculations were used for the \X{}, \Ap{}, \D{} (using A$_2$ symmetry), \Da{}, \Db{} (using A$_2$ symmetry) and \De{} states, while icMRCI(2) calculations (requesting 2 states in the C$_{2v}$ icMRCI) were used for the \A{}, \Dc{} (using A$_2$ symmetry) and \Df{} and icMRCI(3) calculations were used for the \Dd{} state. % (\alert{CHECK THAT THIS IS CORRECT - I MIGHT HAVE USED FEWER IN SOME CASES}). 
For spin-orbit coupling elements, 
the CASSCF is state-averaged over the states required for the bra and ket wavefunction.. 

Following recent recommendations to improve reproducibility of quantum chemistry results,\cite{jt632} we include sample input files in the Supplementary Information. 

\subsection{Results and Discussion}
Full quantitative results are provided in the Supplementary Information in tabular format. 

%\alert{Add some graphs}
\begin{figure}
\includegraphics[width=\textwidth]{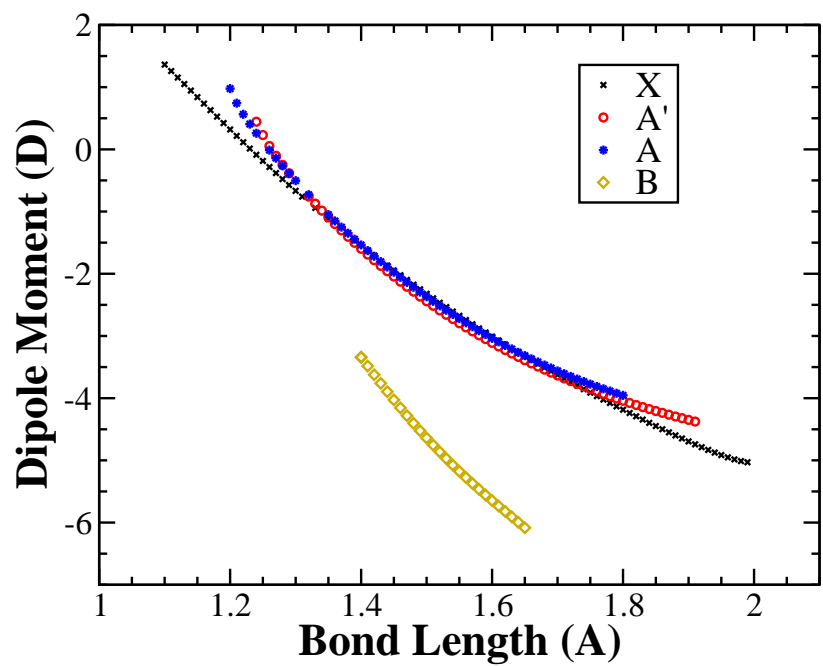}
\caption{\label{fig:DMQuartets} Diagonal dipole moment of quartet states}
\end{figure}

\begin{figure}
\includegraphics[width=\textwidth]{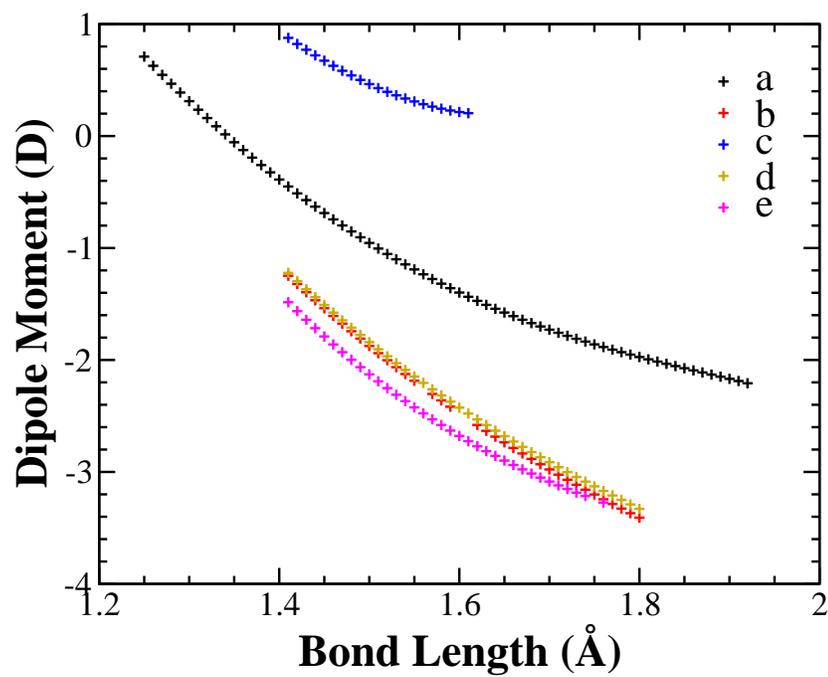}
\caption{\label{fig:DMDoublets} Diagonal dipole moment of doublet states}
\end{figure}

\subsubsection{Diagonal Dipole Moments}
Diagonal dipole moments are only needed for those states that are expected to have significant thermal population at 5000 K; we choose to include the four lowest quartet and five lowest doublet states here. 

\Cref{fig:DMQuartets} and \Cref{fig:DMDoublets} shows the \abinitio\ data for the quartet and doublet states respectively. For all states except the \Da state, it is clear that the electronic state is of largely ionic character due to the near linearity of the dipole moment with increased bond length. The \X, \Ap, \A, \Db, \Dd{} and \De{}  states have similar charge distributions (i.e. ionic separation of charge), while the \B{} state has a much larger asymmetric distribution of electric charge, causing the significant increase in the magnitude of the \B{} state dipole moment. This is consistent with the results of Ref. \cite{07MiMaxx.VO}. On a basic level, the B state has $3d^5$ valence occupancy whilst the others have $4s^13d^4$ valence occupancy. From this, the adverse effects of including the \B{} state in the CASSCF orbital optimisation for the \A{} state are clearly explained. 

%The dipole moment of all these states increases as the bond length increases, consistent with the ionic nature of these states in this bond length region.
Despite the ionic characters of the VO electronic states in the region studied, these states all disassociate to neutral atomic V+O. Therefore, at some point, there is an ionic/covalent avoided crossing. This is the reason why the \abinitio\ results often do not extend beyond about 1.8 \AA. Convergence beyond this region was difficult because the ionic and covalent states were of similar energies. However, the dipole moment obtained is very sensitive to the ionic/covalent character of the state. Therefore, even if calculations converged, the dipole moment obtained was often unpredictable and not smooth in this region. Changes in basis set and/or method etc. did not significantly improve smoothness and convergence and so our methodology was not changed here. 

Details on the way in which these diagonal dipole moments are used in the final line list are given in Ref. \cite{jt644}. In brief, for our final line list, we choose to use \abinitio\ points only where the calculations were trusted and the curves smooth. Using a diabatic-type approach to the ionic/covalent avoided crossing, we derived an appropriate functional form for the shape of the dipole moment going from ionic to covalent character. We then fit these \abinitio\ points to the functional form to obtain smooth curves with the correct physics for our spectroscopic model. 

However, this is unsatisfying from a quantum chemistry perspective. Fundamentally, the CASSCF orbitals struggle to represent both the ionic and covalent structures fairly and consistently. It would be preferable to obtain different orbitals for both states, then use both of these orbitals sets (that may be non-orthogonal) in a subsequent MRCI-type calculation. New multi-reference methodologies as well as approaches based on the old but under-utilised valence-bond orbitals, could help address this problem.

\begin{figure}
\includegraphics[width=\textwidth]{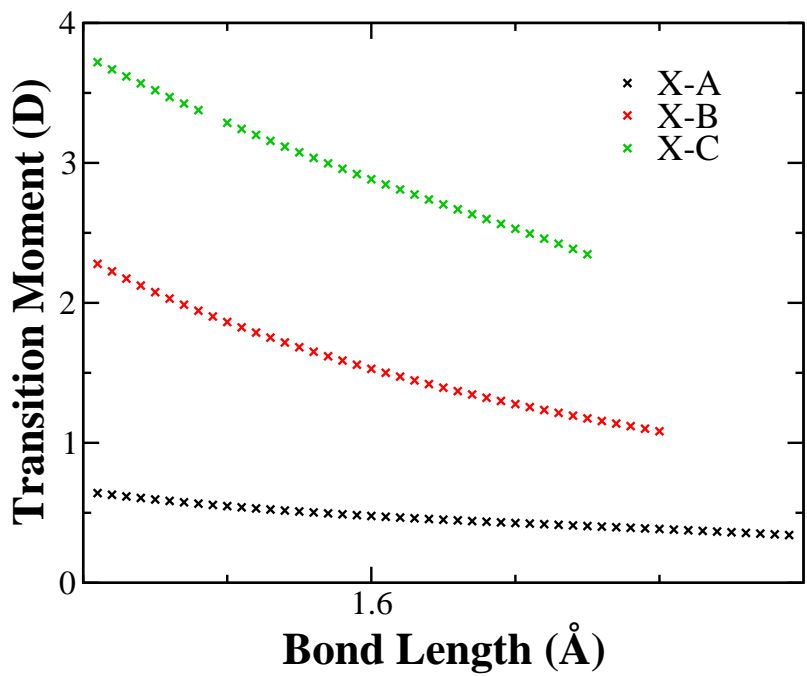}
\caption{\label{fig:OffDiagMain} The main three off-diagonal dipole moments curves for the absorption spectroscopy of VO.}
\end{figure}

\begin{figure*}
\includegraphics[width=\textwidth]{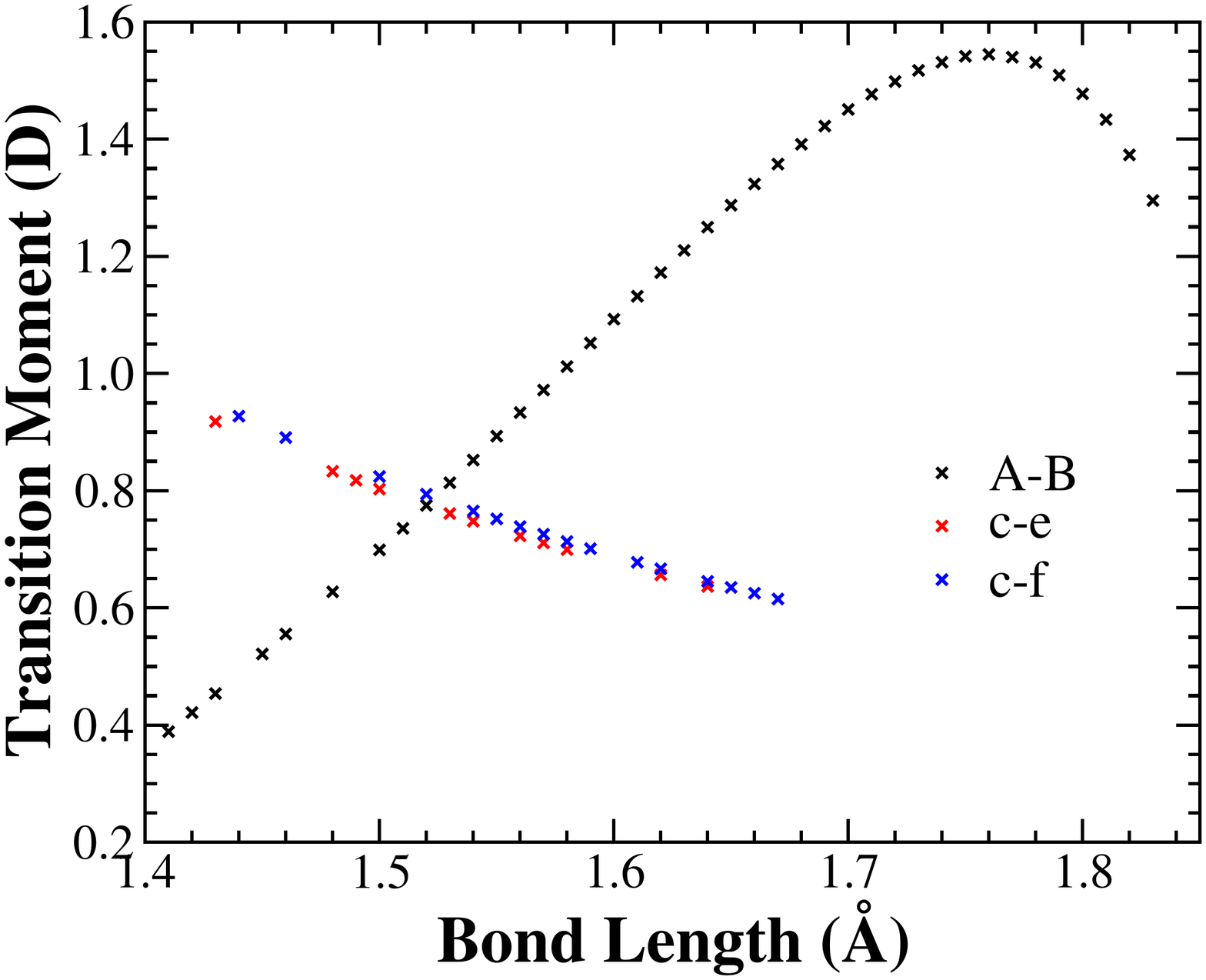}
\includegraphics[width=\textwidth]{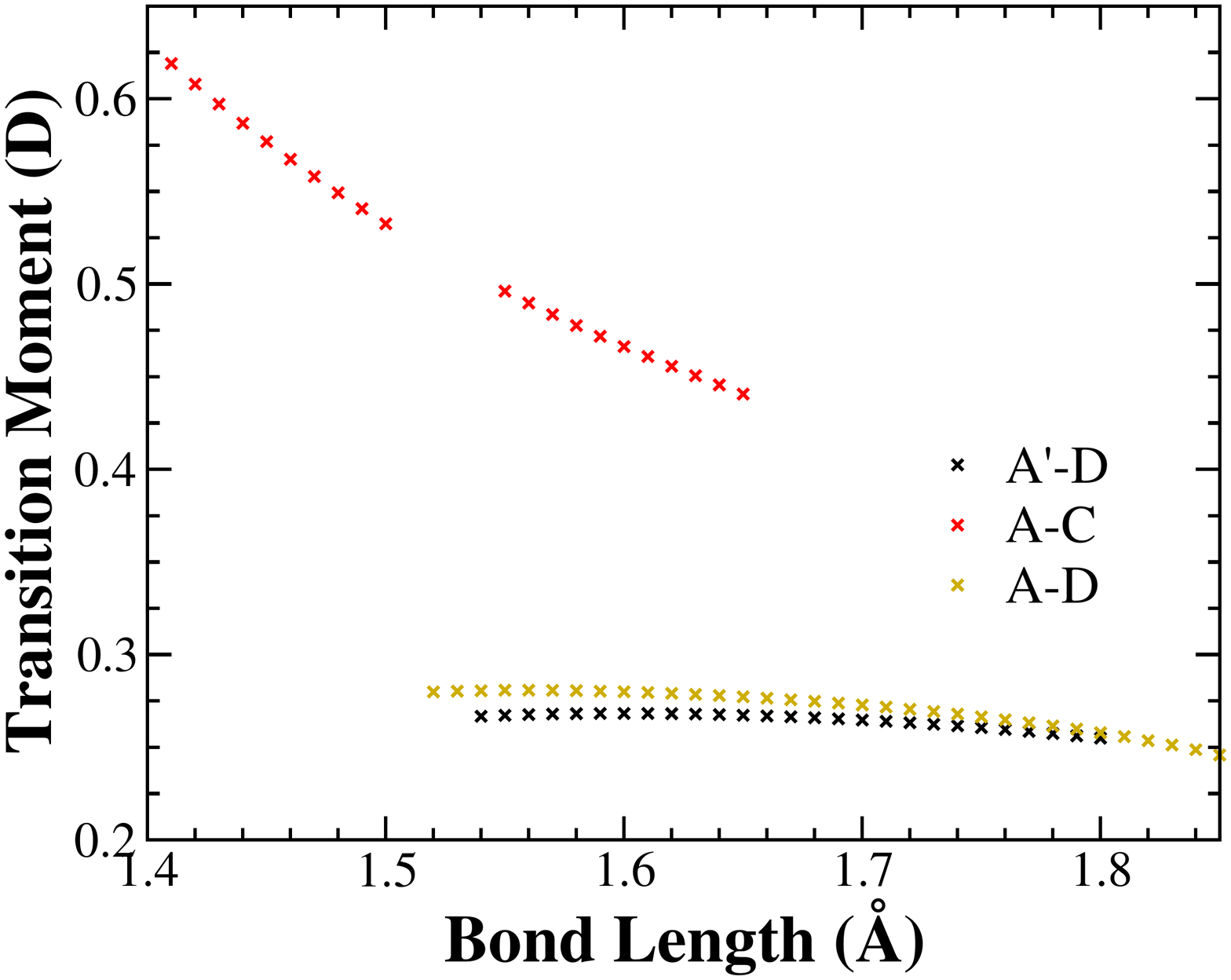}
\caption{\label{fig:OffDiagBigDM} Off-diagonal dipole moment curves.}
\end{figure*}
\begin{figure*}
\includegraphics[width=\textwidth]{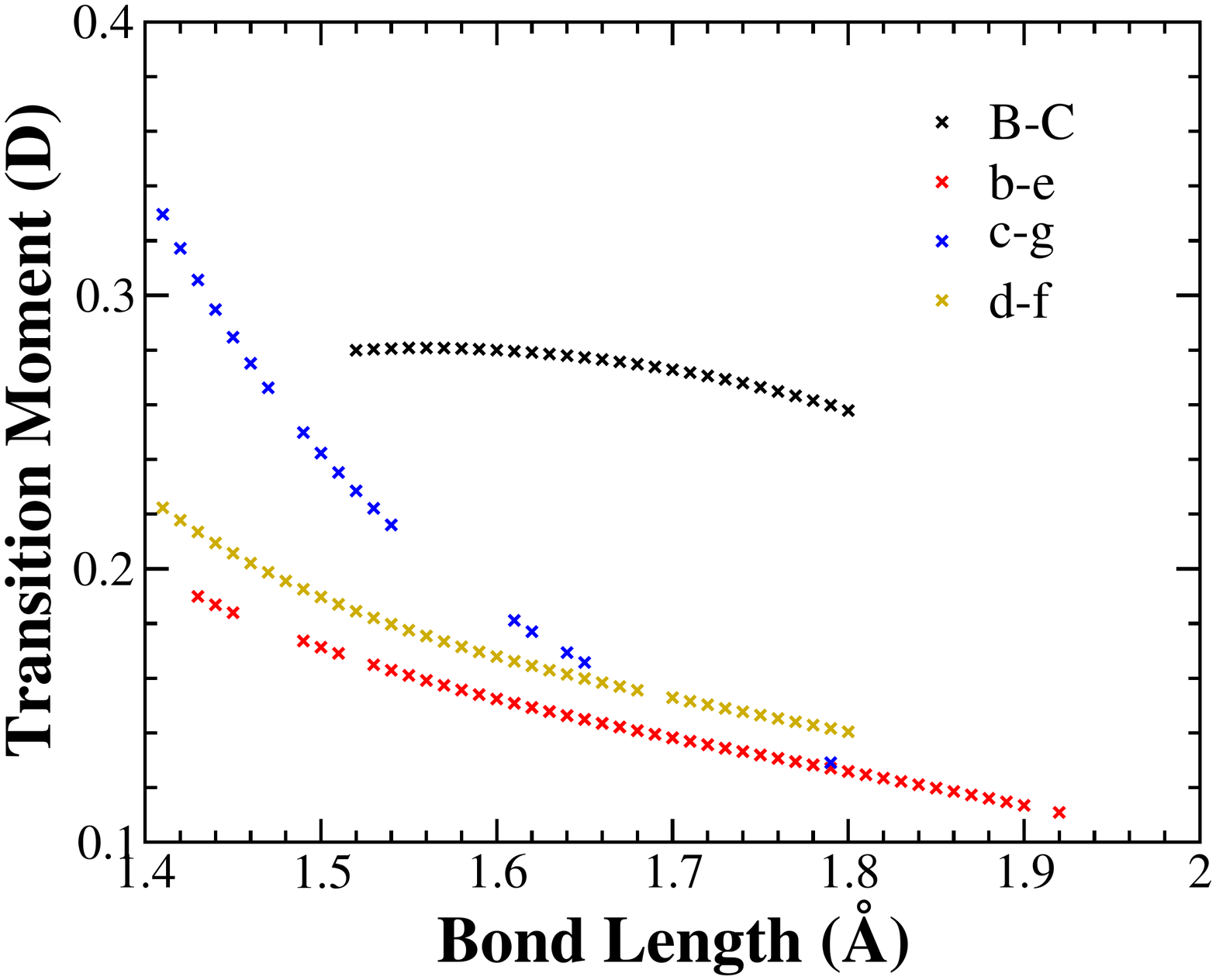}
\includegraphics[width=\textwidth]{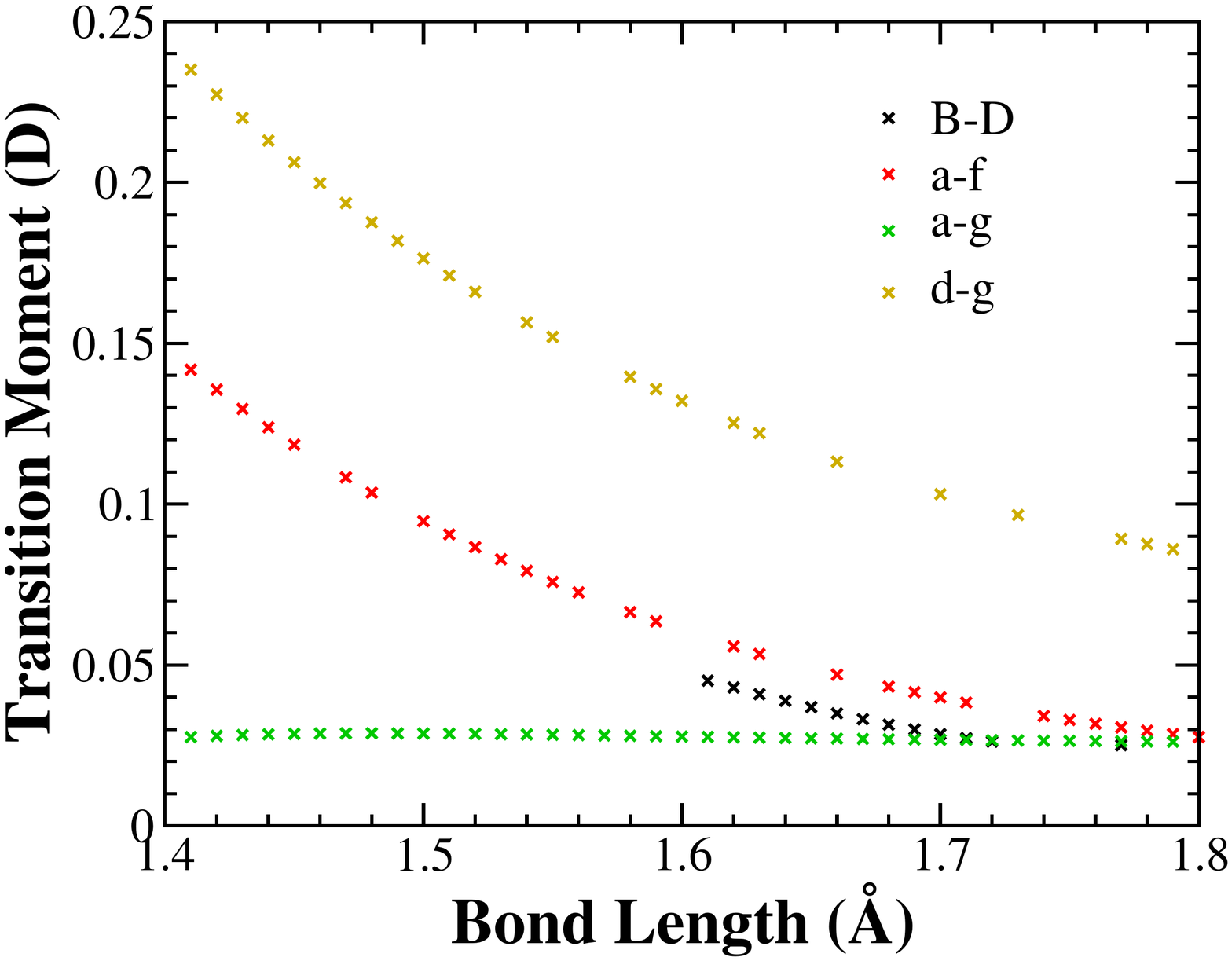}
\caption{\label{fig:OffDiagBigDM2} Off-diagonal dipole moment curves, continued.}
\end{figure*}

\subsubsection{Off-diagonal Dipole Moments}
\Cref{fig:OffDiagMain} shows the final results for the off-diagonal dipole moment for the three main bands in VO absorption spectra. The quality of these curves is the major factor in the quality of the final line list. All three curves are smooth and tend consistently to zero at long bond distances, as required. The C-X transition moment is significantly stronger than the B-X transition moment, which is is significantly stronger than the A-X transition moment. The strength of the final absorption is approximately proportional to the square of this transition moment multiplied by the frequency of the transition and so can be expected to follow this trend for a flat input light source. However, the input light is generally a black-body like; therefore, the importance of absorption in each of these three bands on the atmospheric physics depends on the temperature of the light source. Generally VO is present in the atmospheres M dwarf stars which emit black-body radiation with temperatures around 2000 K. The emission of these objects peaks near the A-X 0-0 transition and therefore this weaker transition will probably have the most influence on the radiative transfer within the M dwarf. Conversely for hot Jupiters illuminated by hotter stars, the black-body peak will move towards the B-X and C-X transition, which will then become more important in the physics of the hot Jupiter atmosphere. All three of these transition moments should be accurate to about 5-10 \%, given our considerations in the previous section.  

\Cref{fig:OffDiagBigDM} and \Cref{fig:OffDiagBigDM2} shows the other off-diagonal dipole moments we consider. These are all transition moments zero by symmetry or spin that both originate from electronic states with term energies less than 16,000 \cm{} (i.e. A', A, B, a, b, c, d, e) and go to one of the lowest 13 electronic states. All curves are relatively smooth as required for our final application; achieving smoothness required careful selection of points in many cases. Most curves go smoothly from a peak towards zero. The main exceptions are the A-B, A'-D, A-D and B-C transitions. The unusual shape of the A-B spectra, however, is of some importance. Other basis sets and method choices did not significantly affect this curve, leading us to conclude that this is probably the true shape of this transition moment. The A'-D, A-D and B-C transition all have similar shapes that are flatter than would be expected; however, these are relatively weak and were not investigated further at this stage. 

The availability of experimental results, particularly on excitation energies and vibrational frequencies, can greatly increase the accuracy of the final spectroscopic model\cite{jt589} and help correct for {\it ab initio} inaccuracies. There are many experimental results  available for VO \cite{82ChHaMe.VO,82ChTaMe.VO,87MeHuCh.VO,91SuFrLo.VO,94ChHaHu.VO,95AdBaBe.VO,97KaLiLu.VO,02RaBeDa.VO,05RaBexx.VO,08FlZixx.VO,09HoHaMa.VO} though more data, particularly for the higher vibrational states, would be very useful.

\begin{figure}
\includegraphics[width=\textwidth]{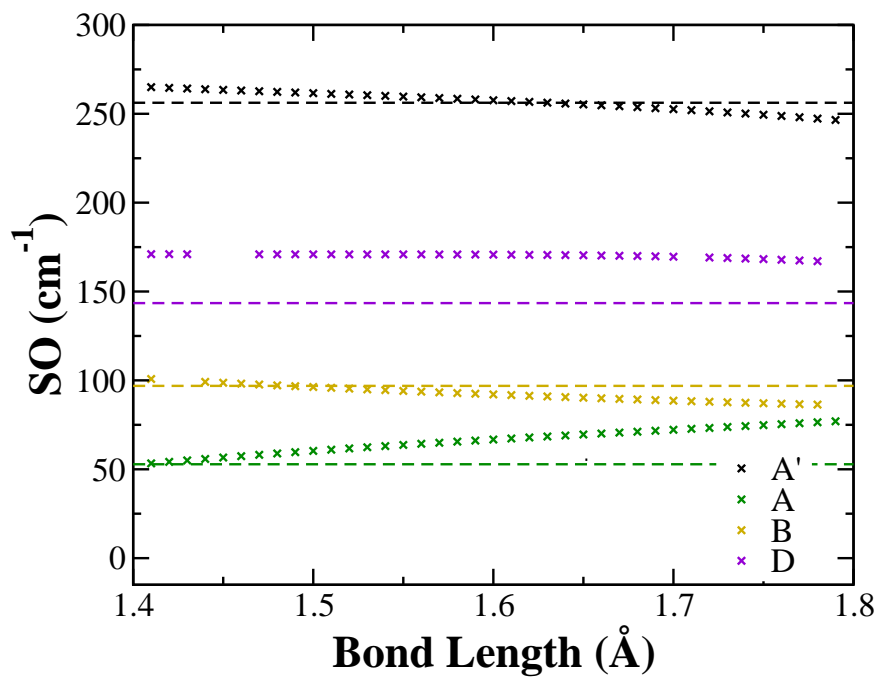}
\includegraphics[width=\textwidth]{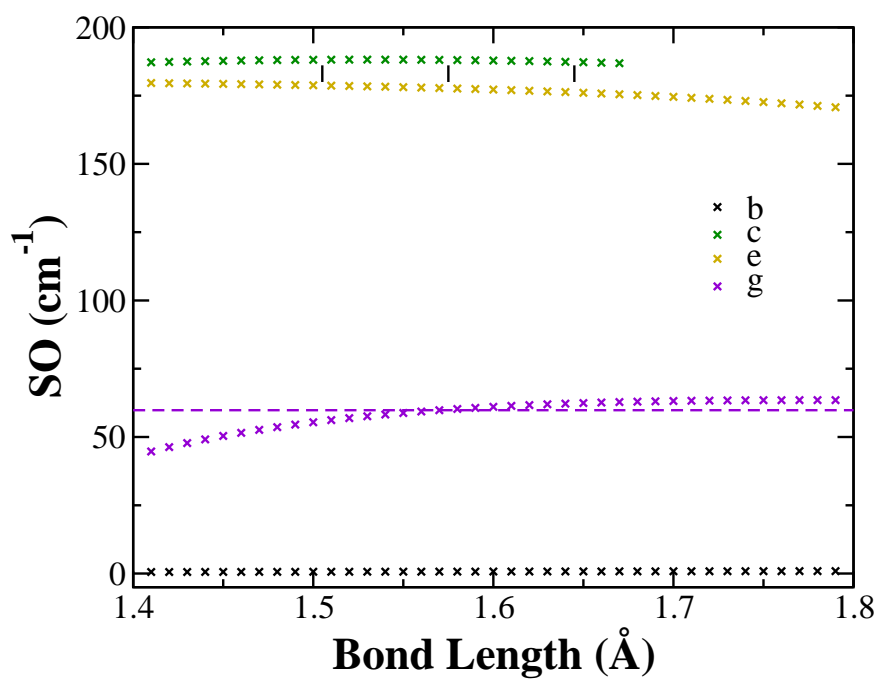}
\caption{\label{fig:SO_DiagQuartet} Diagonal spin-orbit coupling curves. Crosses indicate \abinitio\ data points. Horizontal dotted lines indicate empirical $v=0$ spin-orbit coupling constants from Merer \cite{89Mexxxx.VO}. The black vertical lines in the bottom plot illustrate visually the empirical difference between the \Dc{} and \De{} spin-orbit coupling constant (the absolute value of these constants has not been extracted from experiment).  }
\end{figure}

\subsubsection{Diagonal Spin-Orbit Coupling Curves}% Dipole Moments}
The diagonal spin-orbit coupling curves are shown in \Cref{fig:SO_DiagQuartet}, with experimental equilibrium values when known given by dotted horizontal lines. The agreement between experiment and theory is generally quite high. The variation of this coupling with bond length is reasonably small. The significant differences in the magnitudes of the various SO coupling constants and the relatively high accuracy of the \abinitio\ calculations mean that this property can be used as a key characteristic allowing matching of experimental and theoretical electronic states at energies where the state density is large and the absolute value and relative positioning of energy levels is not sufficiently accurate. % dense regions.  % that th

\begin{figure}
\includegraphics[width=\textwidth]{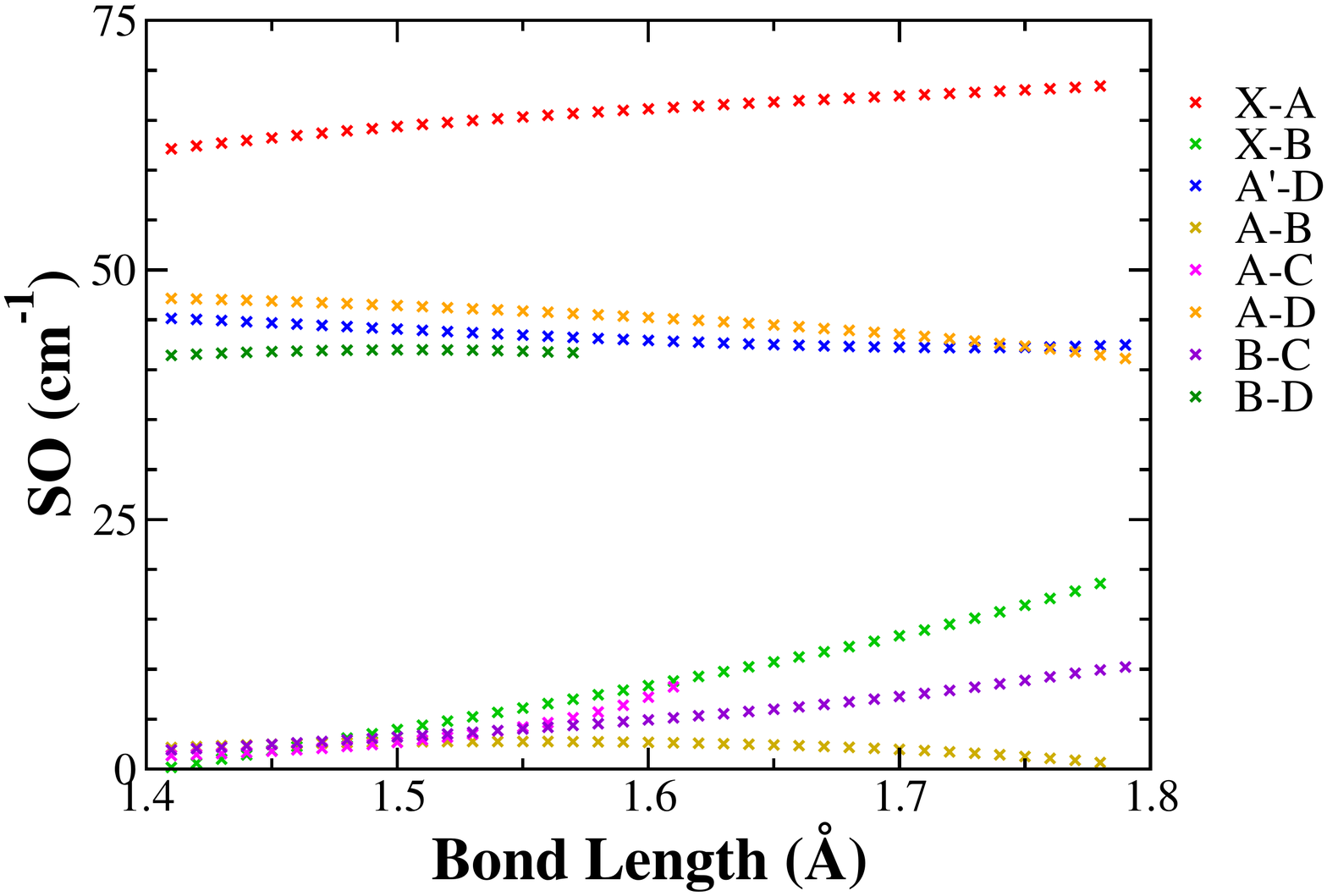}
\includegraphics[width=\textwidth]{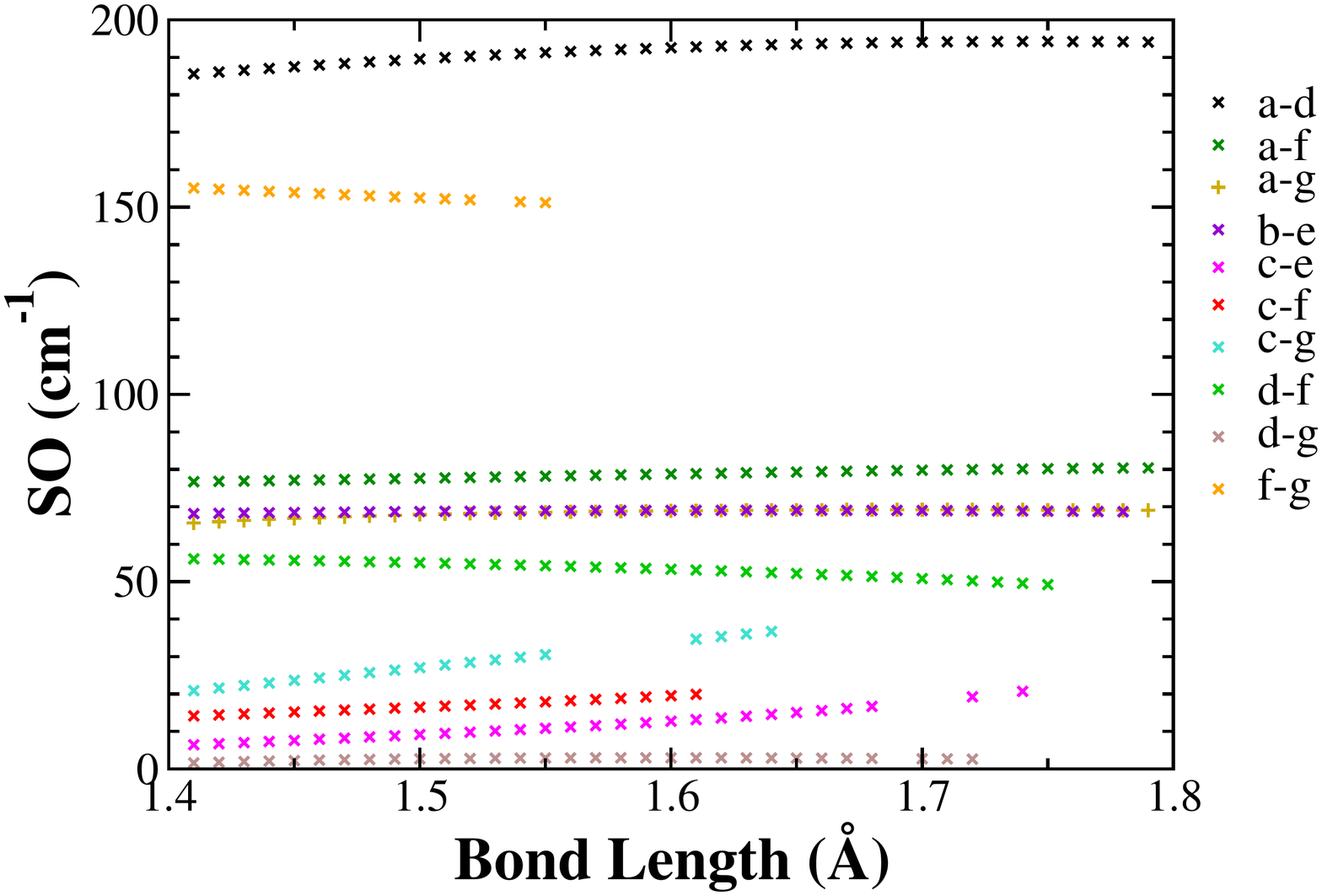}
\caption{\label{SO_OffDiagQuQu2} Off-diagonal spin-orbit coupling curves between states of the same spin.}
\end{figure}

\begin{figure}
\includegraphics[width=0.75\textwidth]{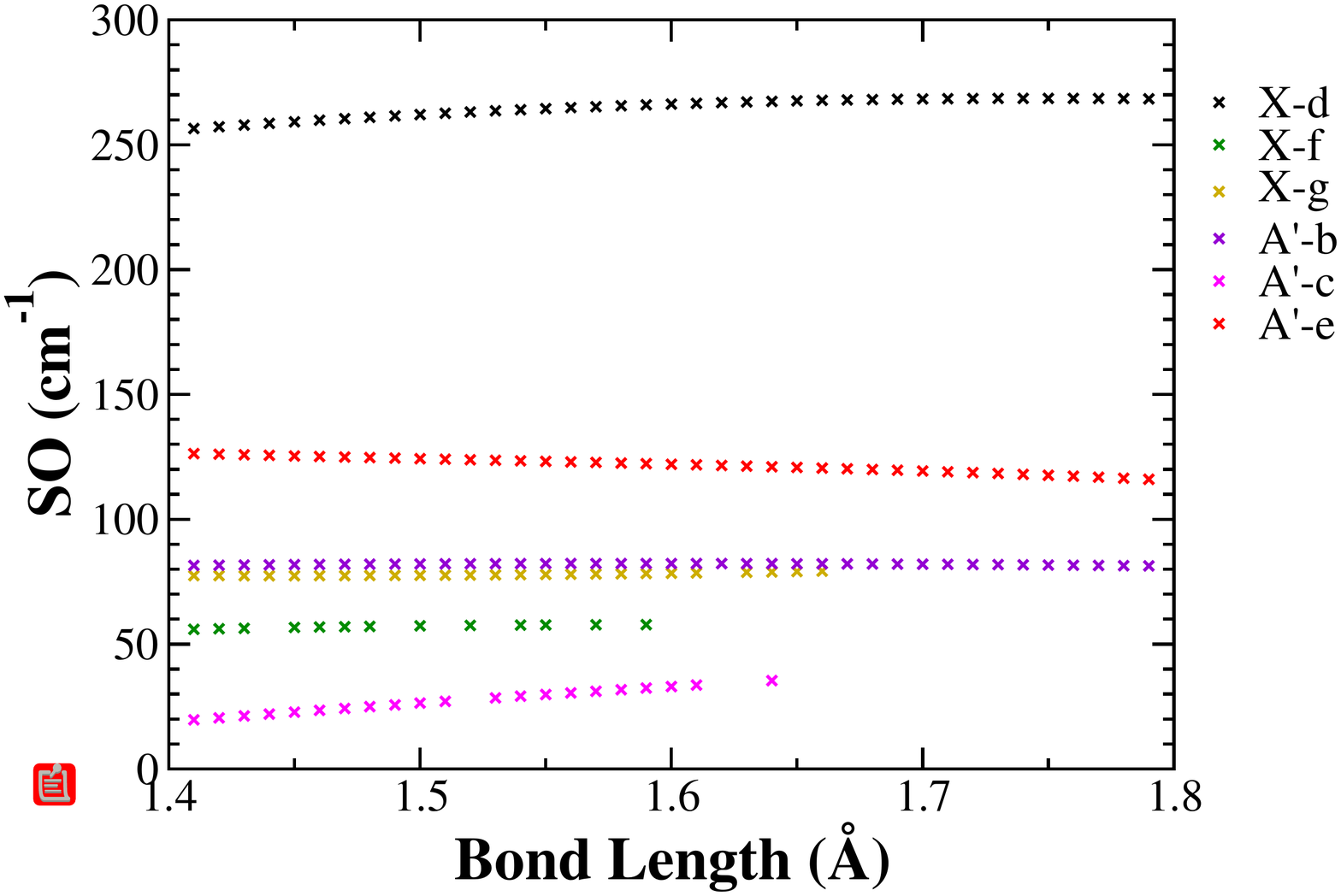}

\includegraphics[width=0.75\textwidth]{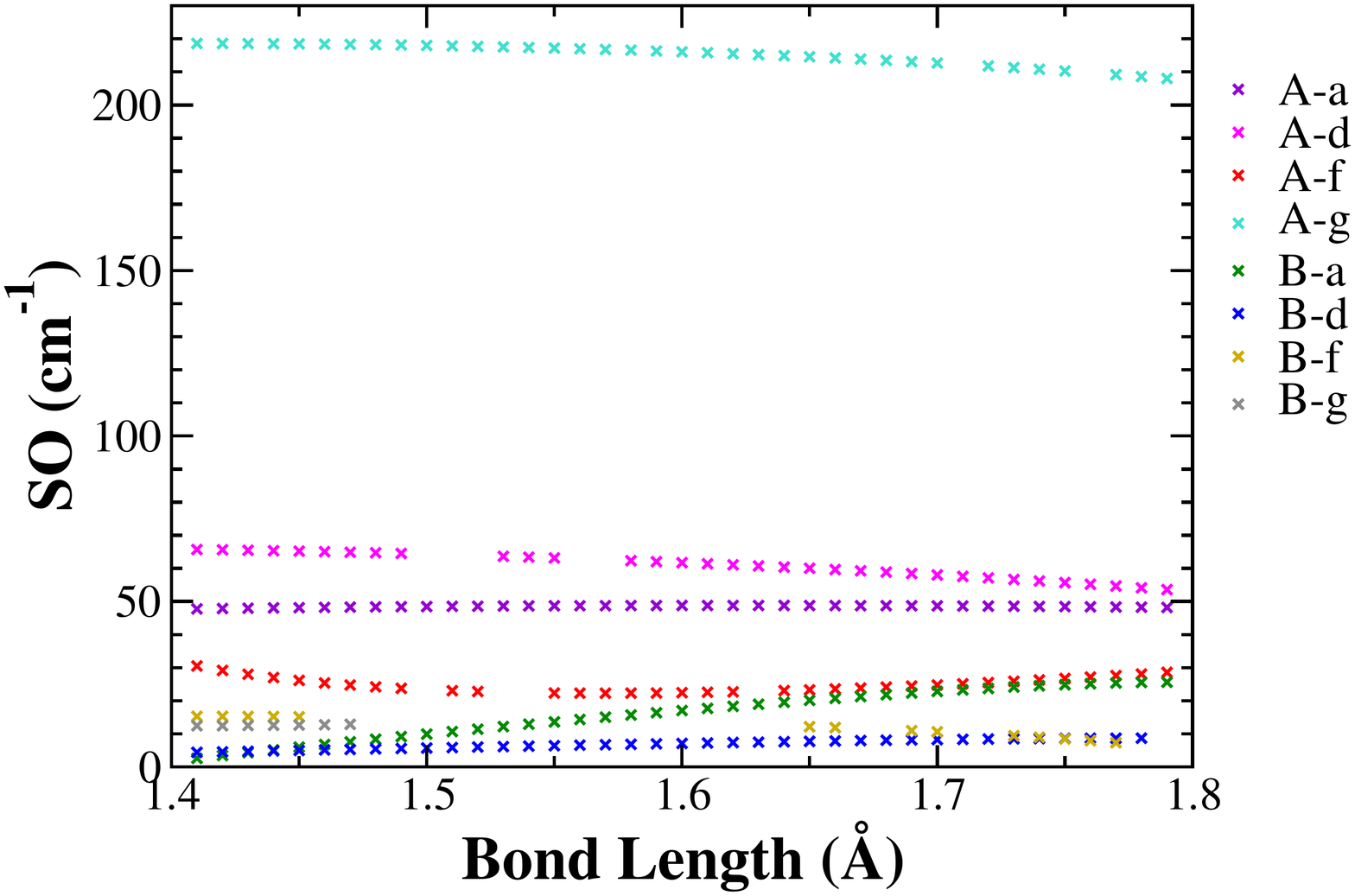}

\includegraphics[width=0.75\textwidth]{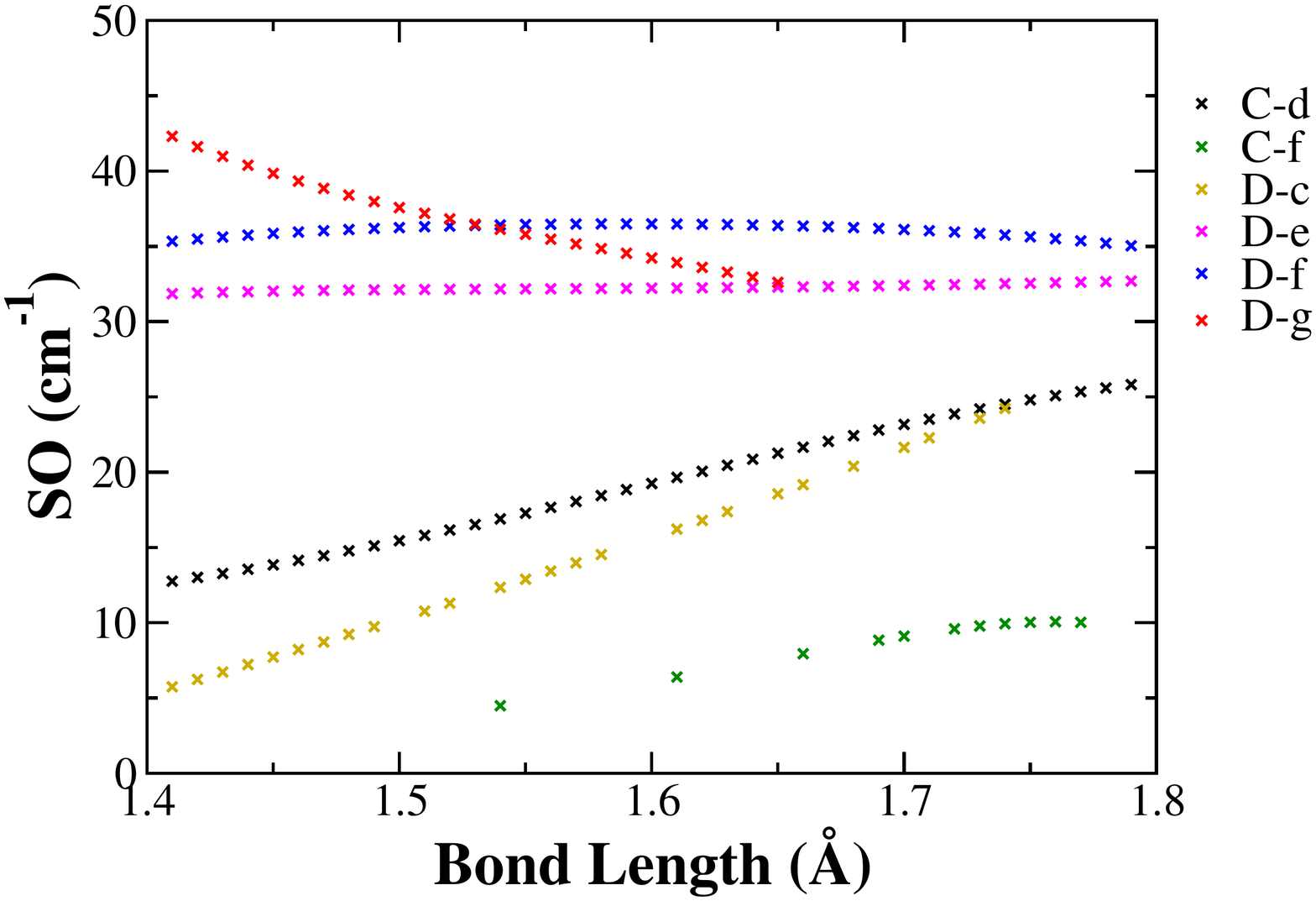}
\caption{\label{SO_OffDiagQD} Off-diagonal spin-orbit coupling curves between states of the different spin.}
\end{figure}

\subsubsection{Off-diagonal Spin-Orbit Coupling Curves}
The off-diagonal spin-orbit coupling between states of the same spin are shown in \Cref{SO_OffDiagQuQu2}. The relative sizes of the couplings can be seen visually. Their effect on the energy levels will depend approximately on the square of the coupling divided by the energy gap. The a-d coupling is particularly strong; however, the influence of this coupling on the absorption spectra of VO is likely to be minimal. The f-g coupling is also quite large, yet there is no experimental evidence of this coupling. This may arise due to the MRCI methodology mixing the \Df{} and \Dg{} states compared to the 'true' answer. A similar argument applies to the A-B spin-orbit coupling, but this is much smaller in magnitude.  % In this case, experiment doesn't support this ------. We hypothesis,  

The off-diagonal spin-orbit couplings between states of different spin are shown in \Cref{SO_OffDiagQD}. These couplings play a very special role in being the only mechanism to allow mixing of electronic states with different spins. This means that  in a VO model that incorporates these coupling terms, the final rovibronic energy levels will not be pure doublet or quartet nature. This gives rise to spin-forbidden transitions. The degree to which these coupling influence energy levels is determined by, approximately, the square of the spin-orbit coupling divided by the energy difference between the two electronic states. % \alert{Need to add influence of $\Lambda$ value if this is relevant??}.
 Therefore, it is clear that the X-d and A-g spin-orbit couplings, being over 200 \cm{}, can cause significant perturbations, mixing and forbidden transitions. This coupling between the X-d state could be the reason why the X-f and X-g transitions can be observed in \cite{09HoHaMa.VO}. Further, the A'-b and A-d spin-orbit couplings are likely to have a significant influence, given the proximity of the two states and the relatively large magnitude of the coupling. When considering the results for the higher electronic states, it is important to note that there are a significant number of doublet states just above the 13 electronic states considered here \cite{15HuHoHi.VO}. The coupling to these states may be as significant as, say, the D-f and D-g couplings quantified in the bottom plot of \Cref{SO_OffDiagQD}. On the scale of \Cref{SO_OffDiagQD}, the off-diagonal spin-orbit couplings seem relatively constant across the bond length; therefore use of an equilibrium value may be appropriate for many comparisons. %a useful approximation in many situations 

\section{Conclusion}

The use {\it ab initio} quantum chemistry for the study of electronic excited states of transition metal diatomics
is definitely not at the `black-box' stage. Results need to be considered carefully and critically; assessing convergence with respect to basis set, orbitals and method can be a useful way of assessing to what
extent results can be trusted in the absence of preferred experimental data.

We provide some of the first results for off-diagonal dipole moments calculated using the FD methodology. Unfortunately, we find significant issues that prevent its use for transition metal diatomic systems currently. Primarily, there is significant ambiguity over which energy difference to use in the finite-field difference formula. Given the large  errors in the excitation energies with current {\it ab initio} methodologies, this issue inhibits accuracy in the finite-field off-diagonal dipole moment. Furthermore, for perpendicular transitions, calculations times are significantly increased, by about an order of magnitude. This is due primarily to the fact that the applied electric field has to be perpendicular to the molecular axis which reduces the symmetry of the system.

These problems with evaluating accurate transition moments for transition metal diatomics fundamentally arises due to the inadequacy of orthogonal one-electron orbitals as a good first order approximation of the electron distribution around the transition metal centre. Using a large active space helps alleviate, but does not fully solve, these issues. Differing occupancies of the $3d$ and $4s$ orbitals radically change the characteristics of these orbitals. Thus, ideally, different $3d$ and $4s$ occupancies should be represented by different $3d$ and $4s$ orbitals. However, orbitals optimised for each occupancies would be non-orthogonal to each other, which cannot currently be handled in main-stream quantum chemistry packages. Nevertheless, some preliminary calculations based on this reasoning have been performed. Promising results have recently been found in atomic systems using multi-configuration methods with non-orthogonal orbitals \cite{95OlGoJo,05ToHixx.Mn,11Tayal.Fe}. 
%\alert{Can't find a reference for this statement - Jonathan suggested Petsalakis, ID (Petsalakis, Ioannis D.); Theodorakopoulos, G (Theodorakopoulos, Giannoula) but their work that I have found seems to be on molecules and around 1985 - i.e. not recent advances. There is some work on non-orthogonal configuration interaction, but again in molecules. I can't find the atomic literature on this topic}.
A similar approach was recently tested for \ce{Cr$_2$} by Olsen \cite{15Olxxxx.ai}; we welcome further development on this methodology and hope in particular for more accurate electronic excitation energies, which may allow finite-field difference methodology for off-diagonal dipole moments to be sufficiently reliable and robust to be useful.

Despite this gloomy picture, there are some things that are well represented by \emph{ab initio} methodology. Spin-orbit coupling constants are often surprisingly accurate, even with low level theory, as long as the electronic state is correctly identified. 
The spin-orbit coupling constants are a great way to relate an experimentally observed state to a theoretical prediction, or to identify the same state in different theoretical calculations. 

This work produces recommended \abinitio\ data points for the spectroscopic study of VO, which are given in full in the Supplementary Information. These will be used to fit a full spectroscopic model for VO and, from this model, a full spectroscopic line list for this system. Work on this is well advanced and the results will be reported elsewhere \cite{jt644}.

 Elsewhere \cite{jt644}, we report the refinement of this data to produce a final spectroscopic model for VO, the evaluation of accurate variational solutions using the newly published flexible, nuclear motion
code called {\sc Duo} \cite{jt609}, 
 and a line list (list of rovibronic energy levels and the intensities of transitions between these levels). 
 
\section*{Acknowledgements}

We thank Lorenzo Lodi and Maire Gorman for helpful discussions on this topic. We also thank Jeffrey Reimers for his advice in manuscript preparation. 
This work was supported by the ERC under the Advanced Investigator Project
267219.  The authors acknowledge the use of the UCL Legion High Performance Computing Facility (Legion@UCL), and associated support services, in the completion of this work.

\end{document}